\documentclass[10pt]{article}
\usepackage[english]{babel}
\usepackage{amsmath,amsfonts,amssymb,bm}\interdisplaylinepenalty=2500
\usepackage{longtable}

 \mathchardef\epsilon="0122   \mathchardef\varepsilon="010F
 \mathchardef\theta="123      \mathchardef\vartheta="0112
 \mathchardef\rho="125        \mathchardef\varrho="011A
 \mathchardef\phi="127        \mathchardef\varphi="011E
\ifx\undefined\degrees\def\degrees{\ensuremath{^{\circ}}}\fi
\ifx\undefined\celsius\def\celsius{\ensuremath{^{\circ}\mathrm{C}}}\fi
\ifx\undefined\unit\def\unit#1{\ensuremath{\mathrm{\,#1}}}\fi
\ifx\undefined\micro\def\micro{\ensuremath{\mu}}\fi
\ifx\undefined\sups\def\sups#1{\ensuremath{^{\mathrm{#1}}}}\fi
\ifx\undefined\subs\def\subs#1{\ensuremath{_{\mathrm{#1}}}}\fi
\ifx\undefined\ohm\def\ohm{\ensuremath{\mathrm{\Omega}}}\fi
\def\req#1{(\ref{#1})}
\ifx\pdfoutput\undefined
  \usepackage{graphicx,color}   
  \usepackage[nativepdf,backref,bookmarks]{hyperref} 
  \usepackage{rotating}
\else
  \usepackage[backref,bookmarks]{hyperref}  
  \usepackage[pdftex]{graphicx,color}     
  \usepackage[pdftex]{rotating}              
\fi
\graphicspath{{figs/}}
\def\PrintGraphicFileNeme{0}
\newcommand{\namedgraphics}[2]{
	\parbox{\textwidth}{%
	\ifnum\PrintGraphicFileNeme>0\rotatebox{90}{~\ttfamily\scriptsize#2}\fi%
	\hspace*{\fill}\includegraphics[scale=#1]{#2}\hspace*{\fill}}}
\newcommand{\includefig}[2]{
	\parbox{\textwidth}{%
	 \ifnum\PrintGraphicFileNeme>0\rotatebox{90}{~\ttfamily\scriptsize#2}\fi%
	\hspace*{\fill}\scalebox{#1}{
	\ifx\pdfoutput\undefined\input{\includefigpath#2.pstex_t}
	\else\input{\includefigpath#2.pdftex_t}\fi}\hspace*{\fill}}}
\newcommand{\comment}[1]{\footnote{#1}}
\renewcommand{\comment}[1]{}

\title{The Measurement of AM noise of Oscillators}
\author{Enrico Rubiola\\
\small web page \texttt{http://rubiola.org}
\\[4em]\includegraphics[width=0.35\textwidth]{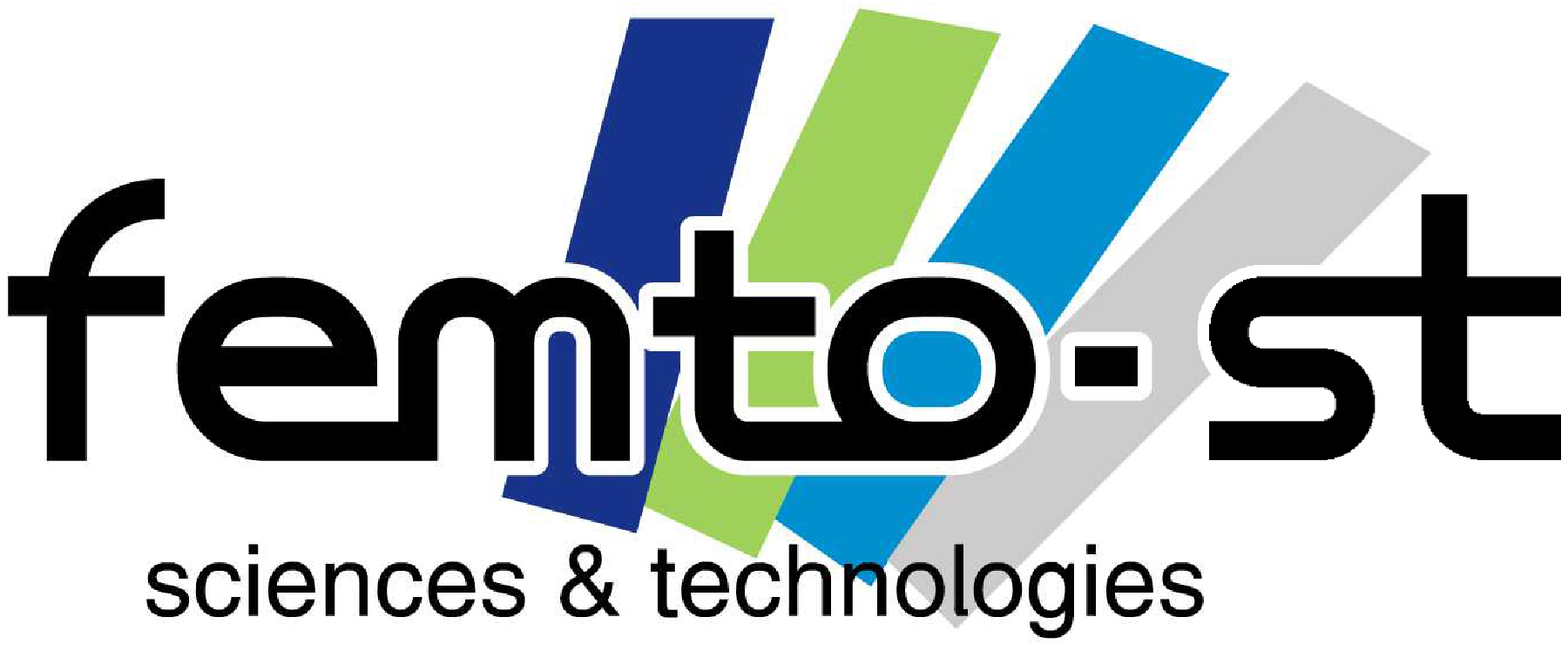}\\[0.5em]
\small FEMTO-ST Institute\\[-0.5ex]
\small CNRS and Universit\'e de Franche Comt\'e, 
\small Besan\c{c}on, France\\[1.5em]}
\date{\small\today}
\begin{document}
\maketitle
\begin{abstract}
\noindent
The close-in AM noise is often neglected, under the assumption that it is a minor problem as compared to phase noise.  With the progress of technology and of experimental science, this assumption is no longer true.  Yet, information in the literature is scarce or absent.

This report describes the measurement of the AM noise of rf/microwave
sources in terms of $S_\alpha(f)$, i.e., the power spectrum density of the
fractional amplitude fluctuation $\alpha$.  The proposed schemes make
use of commercial power detectors based on Schottky and tunnel diodes,
in single-channel and correlation configuration.  

There follow the analysis of the front-end amplifier at the detector output, the analysis of the methods for the measurement of the power-detector noise, and a digression about the calibration procedures.

The measurement methods are extended to the relative intensity noise (RIN) of optical beams, and to the AM noise of the rf/microwave modulation in photonic systems.

Some rf/microwave synthesizers and oscillators have been measured, using correlation and moderate averaging.  As an example, the flicker noise of a low-noise quartz oscillator (Wenzel 501-04623E) is $S_\alpha=1.15{\times}10^{-13}/f$, which is equivalent to an Allan deviation of $\sigma_\alpha=4{\times}10^{-7}$.
The measurement systems described exhibit the world-record lowest background noise.

\end{abstract}
\clearpage

\begin{center}
\addcontentsline{toc}{section}{Symbol list}
\begin{longtable}{ll}\hline
\multicolumn{2}{l}{\textbf{\large\rule[-1ex]{0pt}{3.5ex}Symbol list}}\\\hline
\rule{0pt}{2.5ex}%
$\overline{~\rule{0ex}{1ex}~}$\,, as in $\overline{v}(t)$  
		& average, or dc component [of the signal $v(t)$]\\
$\tilde{~}$\,, as in $\tilde{v}(t)$  & ac component [of the signal $v(t)$]\\
$B$		& system bandwidth\\
$\mathbb{E}\{\cdot\}$ & statistical expectation\\
$f$		& Fourier frequency (near dc)\\
$g$		& voltage gain (thus, the power gain is $g^2$)\\
$h$		& $h=6.626{\times}10^{-34}$ J/Hz, Planck constant\\
$h_i$	& coefficient of the power-law representation of noise,\\
		& $S(f)=\sum_ih_if^i$ \\
$i(t)$, $I$	& current, dc current\\
$I(t)$, $I_0$	& optical intensity 
		(Sections \ref{sec:am-optical-systems}--\ref{sec:am-microwave-photonic})\\
$k$			& $1.38{\times}10{-23}$ J/K, Boltzmann constant\\
$k_d$	& detector gain, V/W\@.  Also $k_a$, $k_b$, $k_c$\\
$m$		& in cross-spectrum measurements, no.\ of averaged spectra\\
$m$		& modulation index
		(Sections \ref{sec:am-optical-systems}--\ref{sec:am-microwave-photonic})\\
$P$, $P_0$         & carrier power.  Also $P_a$, $P_b$, $P_m$, etc.\\
$q$			& $q=1.602{\times}10^{-19}$ C, electron charge\\
$R$		& resistance\\
$R_0$		  & characteristic resistance.  Often $R_0=50$ \ohm\\
rms		& root mean square value\\
$S(f)$, $S_x(f)$ & single-sided power spectrum density (of the quantity $x$)\\
$t$                & time\\
$T$, $T_0$	& absolute temperature, reference temperature ($T_0=290$ K)\\
$v(t)$		& (voltage) signal, as a funtion of time\\
$V_0$		& peak carrier voltage (not accounting for noise)\\
$V_T$		& $V_T=\frac{kT}{q}$ thermal voltage. $V_T\approx25.6$ mV at 25 \celsius\\
$\alpha(t)$        & fractional amplitude fluctuation\\
$\delta$, as in $\delta x$  & random fluctuation (of the quantity $x$)\\
$\Delta$, as in $\Delta x$  & deterministic fluctuation (of the quantity $x$)\\
$\eta$		& diode technical parameter [Eq~.\req{eqn:am-define-rd}], $\eta\in[1..2]$\\
$\eta$		& photodetector quantum efficiency 
			(Sections \ref{sec:am-optical-systems}--\ref{sec:am-microwave-photonic})\\
$\lambda$		& wavelength\\
$\lambda$, as in $\nu_\lambda$	& in a subscript, $\lambda$ means `light' (as opposed to `microwave')\\
$\mu$, as in $\nu_\mu$	& in a subscript, $\mu$ means `microwave' (as opposed to `light')\\
$\nu$, $\nu_0$	& frequency, carrier frequency\\
$\rho$		& photodetector responsivity, A/W (Sections \ref{sec:am-optical-systems}--\ref{sec:am-microwave-photonic})\\
$\sigma_x(\tau)$   & Allan deviation of the quantity $x$\\
$\tau$, as in $\sigma_x(\tau)$ &  measurement time\\
\rule[-1ex]{0pt}{0ex}%
$\phi(t)$          & phase fluctuation\\\hline
\end{longtable}
\end{center}
\clearpage

\tableofcontents


\clearpage

\section{Basics}\label{sec:am-basics}
A quasi-perfect rf/microwave sinusoidal signal can be written as
\begin{equation}
v(t) = V_0\bigl[1+\alpha(t)\bigr]\cos\bigl[2\pi\nu_0 t+\phi(t)\bigr]~~,  
\label{eqn:am-def-alpha}
\end{equation}
where $\alpha(t)$ is the fractional amplitude fluctuation, and
$\phi(t)$ is the phase fluctuation.  Equation \req{eqn:am-def-alpha}
defines $\alpha(t)$ and $\phi(t)$.  In low noise conditions, that is, 
$|\alpha(t)|\ll1$ and $|\phi(t)|\ll1$, Eq.~\req{eqn:am-def-alpha} is 
equivalent to 
\begin{gather}
v(t) = V_0\cos(2\pi\nu_0t) + v_c(t)\cos(2\pi\nu_0t) - v_s(t)\sin(2\pi\nu_0t)\\
\text{with}\qquad\alpha(t)=\frac{1}{V_0}\,v_c(t) 
\qquad\text{and}\qquad
\phi(t)=\frac{1}{V_0}\,v_s(t) \nonumber~.
\end{gather}

We make the following assumptions about $v(t)$, in agreement with actual cases of interest:
\begin{enumerate}\label{list:am-conditions}
\item The expectation of the amplitude is $V_0$.  Thus
  $\mathbb{E}\{\alpha(t)\}=0$.  
\item The expectation of the frequency is $\nu_0$.  Thus
  $\mathbb{E}\{\dot{\phi}(t)\}=0$. 
\item Low noise. $|\alpha(t)|\ll1$ and $|\phi(t)|\ll1$.
\item Narrow band.  The bandwidth of $\alpha$ and $\phi$ is
  $B_\alpha\ll\nu_0$ and $B_\phi\ll\nu_0$.
\end{enumerate}
It is often convenient to describe the close-in noise in terms of the single-side\footnote{Most experimentalists prefer the single-side power spectrum density because all instruments work in this way.  This is because the power can be calculated as $P=\int_0^BS(f)df$, which is far more straightforward than integrating over positive and (to some extent, misterious) negative frequencies.} power spectrum density $S(f)$, as a function of the Fourier frequency $f$.  A model that has been found useful to describe $S(f)$ is the power-law $S(f)=\sum_ih_if^i$.  In the case of
amplitiude noise, generally the spectrum contains only the white noise
$h_0f^0$, the flicker noise $h_{-1}f^{-1}$, and the random walk
$h_{-2}f^{-2}$.  Accordingly,
\begin{equation}
S_\alpha(f)=h_0+h_{-1}f^{-1}+h_{-2}f^{-2}~~.
\label{eqn:am-microw-salpha}
\end{equation}
Random walk and higher-slope phenomena, like drift,
are often induced by the environment.  It is up to the experimentalist to judge the effect of environment.

The spectrum density can be converted into Allan variance using the
formulae of Table~\ref{tab:am-allan-variance}.
\begin{table}
\caption{Relationships between power spectrum density
 and Allan variance.}
\label{tab:am-allan-variance}
\begin{center}
\begin{tabular}{|c|c|c|c|}\hline\hline
\rule[-1.5ex]{0ex}{4ex}
noise type    & Spectrum density $S_\alpha(f)$ 
              & Allan variance $\sigma_\alpha^2(\tau)$\\\hline
\rule[-2.5ex]{0ex}{6.5ex}
white  & $h_0$               & $\displaystyle\frac{h_0}{2\tau}$\\
\rule[-1.5ex]{0ex}{4ex}
flicker & $h_{-1}f^{-1}$ & $\displaystyle h_{-1}\;2\ln(2)$\\
\rule[-2.5ex]{0ex}{6ex}
random walk & $h_{-2}f^{-2}$ & $\displaystyle h_{-2}\;\frac{4\pi^2}{6}\,\tau$\\\hline\hline
\end{tabular}
\end{center}
\end{table}

The signal power is 
\begin{align}
P &= \frac{V_0^2}{2R}\bigl(1+\alpha\bigr)^2\\
\intertext{thus}
P &\simeq \frac{V_0^2}{2R}\bigl(1+2\alpha\bigr)
\qquad\text{because}~\alpha\ll1
\end{align}
It is convenient to rewrite $P$ as $P=P_0+\delta P$, with 
\begin{equation}
P_0=\frac{V_0^2}{2R}
\qquad\text{and}\qquad
\delta P\simeq 2P_0\alpha 
\end{equation}
The amplitude fluctuations are measured through the measurement of the
power fluctuation $\delta P$,
\begin{equation}
\label{eqn:am-p-vs-alpha}
\alpha(t)=\frac{1}{2}\frac{\delta P}{P_0}
\end{equation}
and of its power spectrum density,
\begin{equation}
\label{eqn:am-sp-vs-salpha}
S_\alpha(f)=\frac{1}{4}S_\frac{P}{P_0}(f)
		=\frac{1}{4P_0^2}S_P(f)~.
\end{equation}

The measurement of a two-port device, like an amplifier, is made easy by the availability of the reference signal sent to the device input.  In this case, the bridge (interferometric) method \cite{rubiola02rsi} enables the measurement of amplitude noise and phase noise with outstanding sensitivity.  Yet, the bridge method can not be exploited for the measurement of the AM noise of oscillators, synthesizers and other signal sources.  Other methods are needed, based on power detectors and on suitable signal processing techniques.

\section{Single channel measurement}\label{sec:am-single-channel}
\begin{figure}[t]
\centering\namedgraphics{0.8}{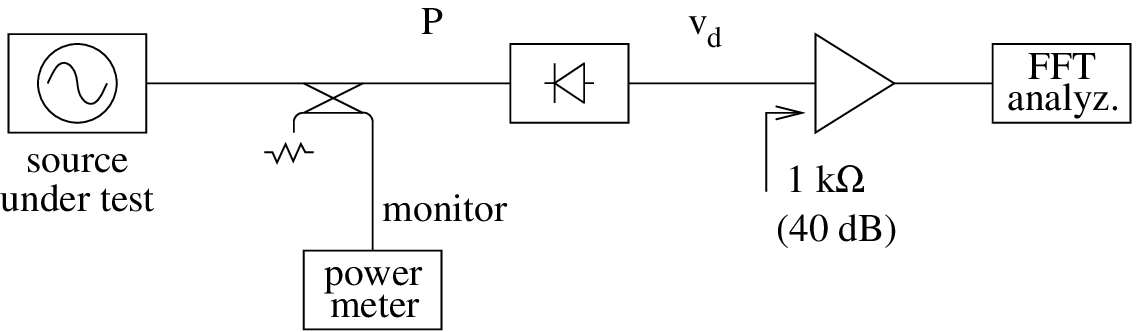}\\[5mm]
\centering\namedgraphics{0.8}{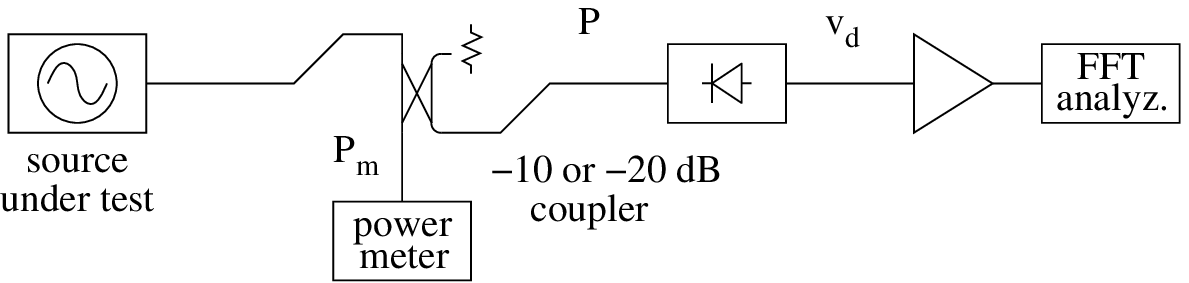}
\caption{Top: Basic scheme for the measurement of AM noise.
  Bottom: A variant useful in some cases where the power detector
  operates at low power, while the power meter does not.}
\label{fig:am-basic-scheme}
\label{fig:am-tricks}
\end{figure}
Figure~\ref{fig:am-basic-scheme} shows the basic scheme for the
measurement of AM noise.  The detector characteristics
(Sec.~\ref{sec:am-detectors}) is $v_d=k_dP$, hence the ac component of
the detected signal is $\tilde{v}_d=k_d\delta P$.  The detected
voltage is related to $\alpha$ by $\tilde{v}_d=k_dP_0\smash{\frac{\delta P}{P_0}}$,
that is,
\begin{equation}
\tilde{v}_d(t)=2k_d P_0\alpha(t)~~.
\end{equation}
Turning voltages into spectra, the above becomes
\begin{equation}
S_{v}(f)=4k_d^2P_0^2S_\alpha(f)~~.
\label{eqn:am-salpha1-1channel}
\end{equation}
Therefore, the spectrum of $\alpha$ can be measured using
\begin{equation}
S_\alpha(f)=\frac{1}{4k_d^2P_0^2}S_{v}(f)~~.
\label{eqn:am-salpha2-1channel}
\end{equation}
Due to linearity of the network that precedes the detector
(directional couplers, cables, etc.), the fractional power fluctuation
$\delta P/P_0$ is the same in all the circuit, thus $\alpha$ is the same. 
As a consequence, the separate measurement of the oscillator power and of
the attenuation from the oscillator to the detector is not necessary.  The straightforward way to use Eq.~\req{eqn:am-salpha1-1channel}, or \req{eqn:am-salpha2-1channel}, is to refer $P_0$ at detector input, and $v_d$ at the detector output.

Interestingly, phase noise has virtually no effect on the
measurement.  This happens because the bandwidth of the detector is
much larger than the maximum frequency of the Fourier analysis, hence
no memory effect takes place.

In single-channel measurements, the background noise can only be assessed by measuring a low-noise source\footnote{The reader familiar with phase noise measurements is used to measure the instrument noise by removing the device under test.  This is not possible in the case of the AM noise of the oscillator.}.  Of course, this measurement gives the total noise of the source and of the instrument, which can not be divided.
The additional hypothesis is therefore required, that the amplitude noise of the source is lower than the instrument background.  Unfortunately, a trusted source will be hardly available in practice. 

Calibration is needed, which consists of the measurement of the
product $k_dP_0$.  See Section~\ref{sec:am-calibration}.

\section{Dual channel (correlation) measurement}\label{sec:am-correlation}
\begin{figure}[t]
\centering\namedgraphics{0.8}{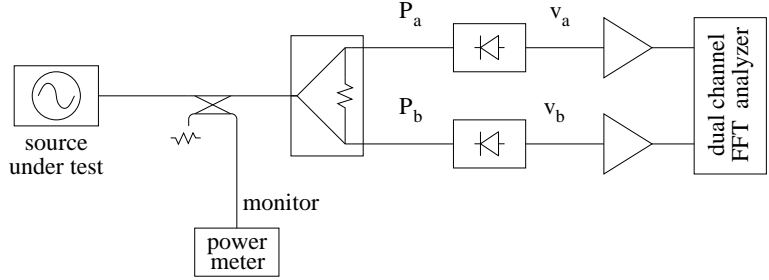}
\caption{Correlation AM noise measurement.}
\label{fig:am-correl-scheme}
\end{figure}
\begin{figure}[t]
\centering\namedgraphics{0.8}{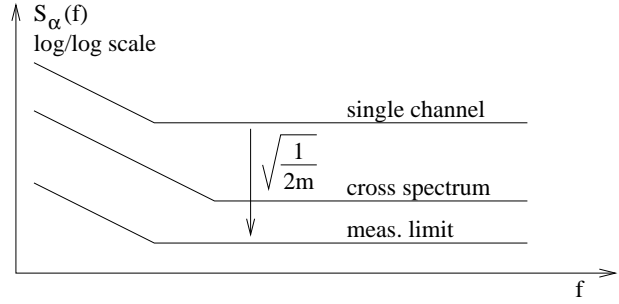}
\caption{Spectra of the correlation AM noise measurement.}
\label{fig:am-spectra}
\end{figure}
Figure~\ref{fig:am-correl-scheme} shows the scheme for the correlation
measurement of AM noise.  The signal is split into two branches, and
measured by two separate power detectors and amplifiers.  Under the assumption that the two channels are independent, the cross
spectrum $S_{ba}(f)$ is proportional to $S_\alpha(f)$.  In fact, the
two dc signals are $v_a=k_aP_a\alpha$ and $v_b=k_bP_b\alpha$.  The
cross spectrum is
\begin{equation}
S_{ba}(f)=4k_ak_bP_aP_bS_\alpha(f)~~,
\label{eqn:am-salpha1-2channel}
\end{equation}
from which
\begin{equation}
S_\alpha(f)=\frac{1}{4k_ak_bP_aP_b}S_{ba}(f)~~.
\label{eqn:am-salpha2-2channel}
\end{equation}
Averaging over $m$ spectra, the noise of the individual channels is
rejected by a factor $\sqrt{2m}$ (Fig.~\ref{fig:am-spectra}), for the
sensitivity can be significantly increased.  A further advantage of
the correlation method is that the measurement of $S_\alpha(f)$ is
validated by the simultaneous measurement of the instrument noise limit, that is, the single-channel noise divided by $\sqrt{2m}$.  This solves one of the major problems of the single-channel measurement, i.e., the need of a trusted low-noise source.

Larger is the power delivered by the source under test, larger is the instrument gain.  This applies to single-channel measurements, where the gain is $4k_d^2P_0^2$ [Eq.~\req{eqn:am-salpha1-1channel}], and to correlation measurements, where the gain is $4k_ak_bP_aP_b$ [Eq.~\req{eqn:am-salpha1-2channel}].  Yet in a correlation system the total power $P_0$ is split into the two channels, for $P_aP_b=\frac{1}{4}P_0^2$.  Hence, switching from single-channel to correlation the gain drops by a factor $\frac{1}{4}$  ($-6$ dB).
Let us now compare a correlation system to a single-channel system under the simplified hypothesis that the background noise referred at the detector output is unchanged.  This happens if the noise of the dc preamplifier is dominant.
In such cases, the background noise referred to the instrument input, thus to $S_\alpha$, is multiplied by a factor $\smash{\frac{4}{\sqrt{2m}}}$.  The numerator ``4'' arises from the reduced gain, while the denominator $\sqrt{2m}$ is due to averaging.  Accordingly, it must be $m>8$ for the correlation scheme to be advantageous in terms of sensitivity.   On the other hand, if the power of the source under test is large enough for the system to work at full gain in both cases, the dual-channel system exhibits higher sensitivity even at $m=1$.

Calibration is about the same as for the single-channel measurements.
See Section~\ref{sec:am-calibration}.

In laboratory practice, the availability of a dual-channel FFT analyzer is the most frequent critical point.  If this instrument is available, the experimentalist will prefer the correlation scheme in virtually all cases.

\section{Schottky and tunnel diode power detectors}\label{sec:am-detectors}
A rf/microwave power detector uses the nonlinear response of a diode
to turn the input power $P$ into a dc voltage $v_d$.  The transfer function is
\begin{equation}
v_d=k_dP~~,
\label{eqn:am-define-kd}
\end{equation}
which defines the detector gain $k_d$.  The physical dimension of $k_d$ is
\unit{A^{-1}}.  The technical unit often used in data sheets is mV/mW, equivalent to \unit{A^{-1}}.
The diodes can only work at low input level.  Beyond a threshold power, the output voltage differs  smoothly from Eq.~\req{eqn:am-define-kd}.  The actual response depends on the diode type.

\begin{figure}[t]
  \centering\namedgraphics{0.75}{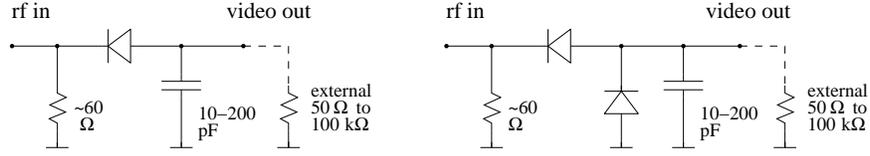}
\caption{Scheme of the diode power detector.}
\label{fig:am-detector-scheme}
\end{figure}
Figure~\ref{fig:am-detector-scheme} shows the scheme of actual power detectors.  The input resistor matches the high input impedance of the
diode network to the standard value $R_0=50$~\ohm\ over the bandwidth
and over the power range.  The value depends on the specific detector.
The output capacitor filters the video\footnote{From the early time of electronics, the term `video' is used (as opposed to `audio') to emphasize the large bandwidth of the demodulated signal, regardless of the real purposes.} signal, eliminating carrier from the output.  A low capacitance makes the detector fast.  On the other hand, a higher capacitance is needed if the detector is used to demodulate a low-frequency carrier. The two-diode configuration provides larger output voltage and some temperature compensation.

\begin{table}[t]
\caption{some power-detector manufacturers (non-exhaustive list).}
\label{tab:am-manufacturers}
\centering
\begin{tabular}[t]{|l|l|}\hline
manufacturer					& web site\\\hline\hline
Aeroflex/Metelics					& aeroflex-metelics.com\\
Agilent Technologies				& agilent.com\\
Advanced Control Components	& advanced-control.com\\
Advanced Microwave				& advancedmicrowaveinc.com\\
Eclipse							& eclipsemicrowave.com\\
Herotek							& herotek.com\\
Microphase						& microphase.com/military/detectors.shtml\\
Omniyig							& omniyig.com\\
RLC Electronics					& rlcelectronics.com/detectors.htm\\
S-Team							& s-team.sk\\\hline\hline
\end{tabular}
\end{table}

Power detectors are available off-the-shelf from numerous manufacturers, some of which are listed on Table~\ref{tab:am-manufacturers}.  Agilent Technologies provides a series of useful application notes \cite{agilent:an-1449} about the measurement of rf/microwave power.

Two types of diode are used in practice, Schottky and tunnel.  Their
typical characteristics are shown in Table~\ref{tab:am-typical-detectors}.
\begin{table}
\caption{Typical characteristics of Schottky and tunnel power detectors.}
\label{tab:am-typical-detectors}
\begin{center}
\begin{tabular}{|l|c|c|}\hline\hline
\rule[-1ex]{0pt}{3.5ex}    & Schottky         &tunnel \\\hline
\rule{0pt}{2.5ex}%
input bandwidth            & up to 4 decades    & 1--3 octaves \\
                                   &10\,MHz to 20\,GHz
                                           & up to 40 GHz\\
\textsc{vsvr} max.\        &  1.5:1           & 3.5:1 \\
max.\ input power (spec.)  & $-15$ dBm        &$-15$ dBm    \\
absolute max.\ input power &   20 dBm or more & 20 dBm \\
output resistance          &   1--10\,k\ohm\  &50--200 \ohm\\
output capacitance         &  20--200 pF      & 10--50 pF \\
gain                       & 300 V/W          & 1000 V/W \\
\rule[-1ex]{0pt}{1ex}%
cryogenic temperature      &     no           & yes\\\hline\hline   
\end{tabular}
\end{center}
\end{table}

\begin{figure}[t]
\namedgraphics{0.8}{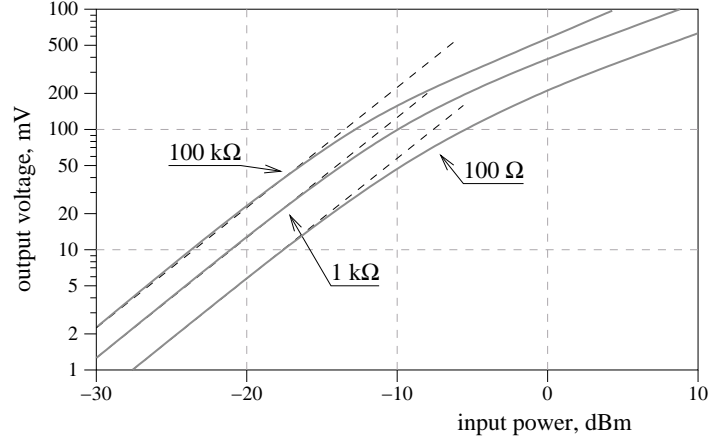}
\caption{Response of a two-diode power detector.}
\label{fig:am-detector-xfer}
\end{figure}

Schottky detectors are the most common ones.  The relatively high
output resistance and capacitance makes the detector suitable to low-frequency carriers, starting from some 10 MHz (typical).  In this condition the current flowing through the
diode is small, and the input matching to $R_0=50$~\ohm\ is provided
by a low value resistor.  Thus, the VSWR is close to 1:1 in a wide
frequency range.  Most of the input power is dissipated in the input
resistance, which reduces the risk of damage in case of overload.  A
strong preference for negative output voltage seems to derive from the
lower noise of P type Schottky diodes, as compared to N type ones, in conjunction with practical issues of mechanical layout. 
Figure~\ref{fig:am-detector-xfer} shows the response of a two-diode Schottky power detector.
The quadratic response [Eq.~\req{eqn:am-define-kd}] derives from the
diode resistance $R_d$, which is related to the saturation current $I_0$ by
\begin{equation}
R_d=\frac{\eta V_T}{qI_0}~~,
\label{eqn:am-define-rd}
\end{equation}
where $\eta\in[1\ldots2]$ is a parameter that derives from the
junction technology; $V_T=kT/q\simeq25.6$~mV at room temperature is
the thermal voltage.  At higher input level, $R_d$ becomes too small
and the detector response turns smoothly from quadratic to linear, like the response of the common AM demodulators and power rectifiers.

\begin{figure}[t]
\namedgraphics{0.5}{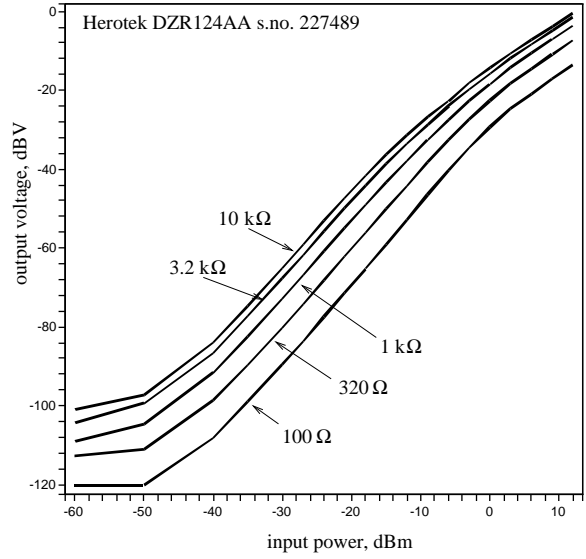}
\caption{Measured response of a Schottky detector Herotek DZR124AA\comment{ (X-110)}.}
\label{fig:am-dzr124aa}
\end{figure}

\begin{figure}[t]
\namedgraphics{0.5}{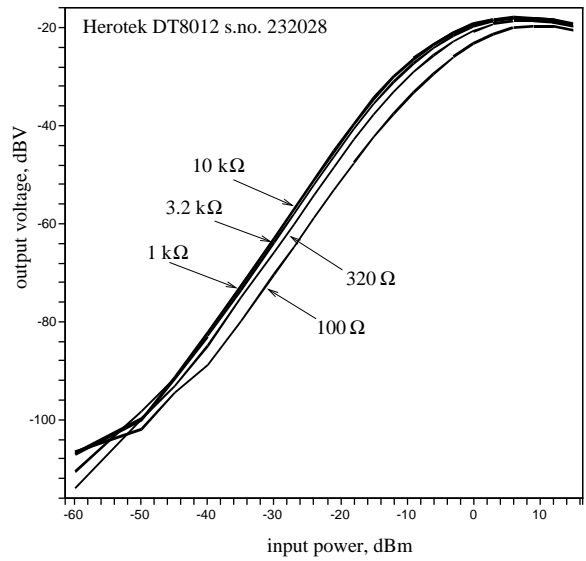}
\caption{Measured response of a tunnel detector Herotek DT8012\comment{ (X-111)}.}
\label{fig:am-dt8012}
\end{figure}

\begin{table}[t]
\caption{Measured conversion gain.}
\label{tab:am-conversion-gain}
\begin{center}
\begin{tabular}{|c|cc|}\hline
			&\multicolumn{2}{|c|}{\rule[0ex]{0ex}{2.5ex}%
			detector gain, $A^{-1}$}\\
load resistance, \ohm	& DZR124AA	& DT8012 \\
			& (Schottky)		& (tunnel)
\comment{\\& s.\,no.\,227489 & s.\,no.\,232028}\\\hline
\rule[0ex]{0ex}{2.5ex}
$1{\times}10^2$		&  35	& 292\\
$3.2{\times}10^2$	&  98	& 505\\
$1{\times}10^3$		&217	& 652\\
$3.2{\times}10^3$	&374	& 724\\
$1{\times}10^4$		&494	& 750
\rule[-1.5ex]{0ex}{1.5ex}\\ \hline
\multicolumn{3}{|c|}{\rule[-1.5ex]{0ex}{4ex}%
	conditions: power $-50$ to $-20$ dBm}\\\hline 
\end{tabular}
\end{center}
\end{table}

Tunnel detectors are actually \emph{backward} detectors.  The backward diode is a tunnel diode in which the negative resistance in the forward-bias region is made negligible by appropriate doping, and used in the reverse-bias region.  Most of the work on such detectors dates back to the sixties \cite{burrus63mtt,gabriel67mtt,hall60ireted}.
Tunnel detectors exhibit fast switching and higher gain than the Schottky counterpart.  A low output resistance is necessary, which affects the input impedance.  Input impedance matching is therefore poor.  
In the measurement of AM noise, as in other applications in which fast response is not relevant, the output resistance can be higher than the recommended value, and limited only by noise considerations.  
At higher output resistance the gain further increases.  
Tunnel diodes also work in cryogenic environment, provided the package tolerates the mechanical stress of the thermal contraction.

Figures~\ref{fig:am-dzr124aa}--\ref{fig:am-dt8012} and Table~\ref{tab:am-conversion-gain} show the conversion gain of two detectors, measured at the FEMTO-ST Institute.
As expected, the Schottky detector leaves smoothly the quadratic law (true power detection) at some $-12$ dBm, where it becomes a peak voltage detector.  The response of the tunnel detector is quadratic up to a maximum power lower than that of the Schottky diode.  This is due to the lower threshold of the tunnel effect.  The output voltage shows a maximum at some 0 dBm, then decreases.  This is ascribed to the tunnel-diode conduction in the forward region.      

At the FEMTO-ST Institute, I routinely use the Herotek DZR124AA (Schottky) and  DT8012 (tunnel).  At the JPL, I have sometimes used the pair
HP432A, and more recently the same Herotek types that I use at the
LPMO.  The old HP432A pair available shows an asymmetry of almost 10
dB in flickering, which causes experimental difficulties.  This asymmetry might be a defect of the specific sample.

\section{The double balanced mixer}\label{sec:am-other-detectors}
%
\begin{figure}[t]
  \centering\namedgraphics{0.8}{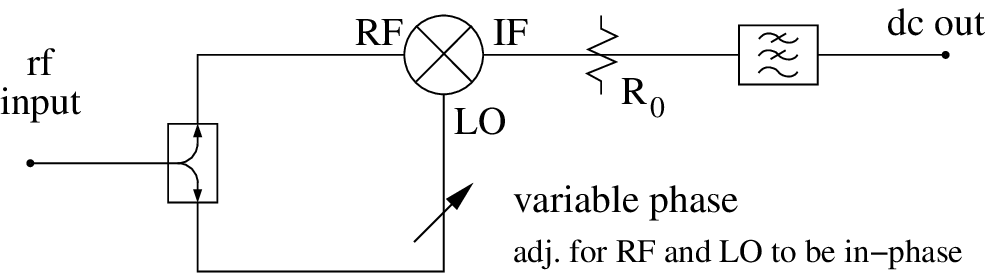}\\[1em]
  \namedgraphics{0.8}{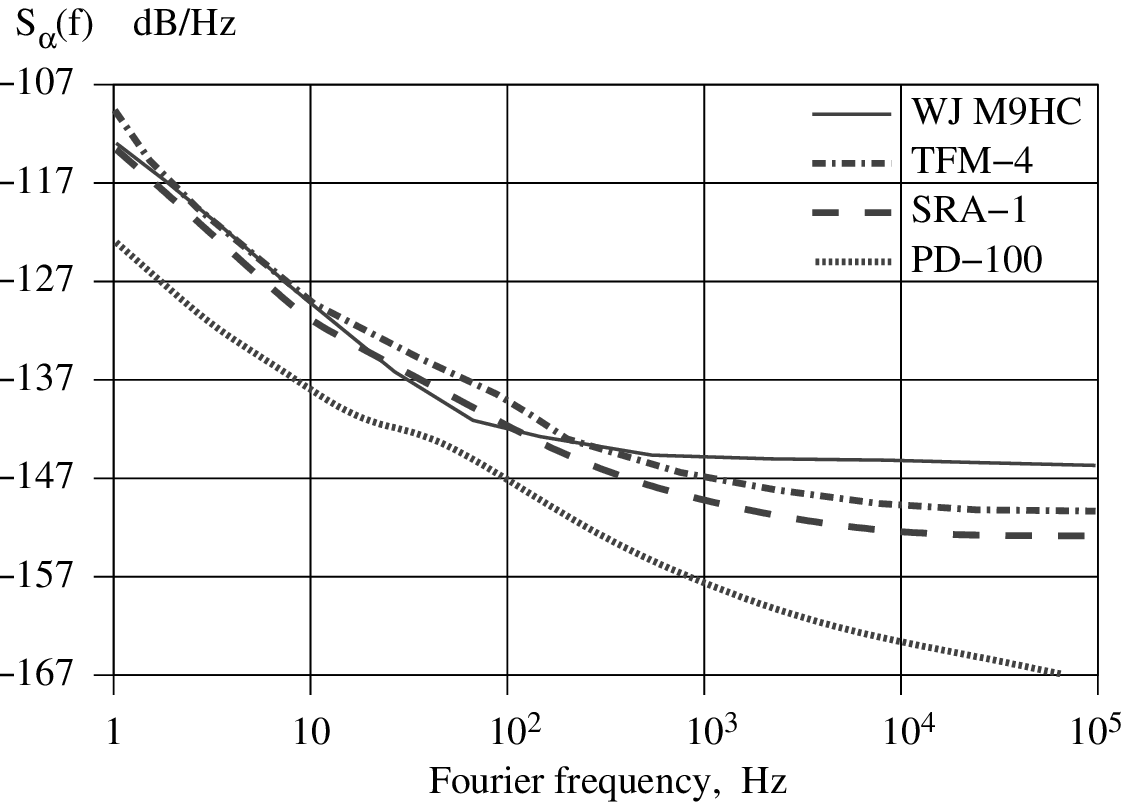}
\caption{A double balanced mixer can be used as the power detector
  for the measurement of AM noise.  Data are from \cite{nelson94uffc}.}
\label{fig:am-mixer}
\end{figure}
Some articles from the NIST \cite{nelson94uffc}\footnote{Fred Walls published some articles in the FCS Proceedings.  Craig Nelson showed the method in his tutorial given at the
2003 FCS.} suggest the use of a double balanced mixer as the power detector in AM noise measurements, with a configuration similar to that of Fig.~\ref{fig:am-mixer}.  Higher sensitivity is obtained with cross-spectrum measurements, using two mixers.
The double balanced mixer is operated in the following conditions:
\begin{enumerate}
\item Both RF and LO inputs are \emph{not} saturated.
\item The RF and LO signals are in-phase.
\end{enumerate}
At low frequencies the in-phase condition may be hard wired, omitting the variable phase shifter.

However useful, the available articles do not explain the diode operation in terms of electrical parameters.  One can expect that the background noise of the mixer used as the AM detector can not be lower than that of a Schottky-diode power detector for the simple reason that the mixer contains a ring of Schottky diodes.  
One could guess that the use of the mixer for AM noise measurements originates from having had mixers and phase shifter on hand for a long time at the NIST department of phase noise.

\section{Power detector noise}\label{sec:am-detector-noise}
Two fundamental types of noise are present in a power detector, shot noise and thermal noise \cite[Sec.~V]{gabriel67mtt}.  In addition, detectors show flicker noise.  The latter is not explained theoretically, for the detector characterization relies on experimental parameters.  Some useful pieces of information are available in \cite{eng61iretmtt}.

Owing to the shot effect, the average current $\overline{\imath}$ flowing in the diode junction is affected by a fluctuation of power spectral density
\begin{align}
S_i=2q\overline{\imath}\qquad\unit{A^2/Hz},
\label{eqn:am-shot-noise}
\end{align}
Using the Ohm law $v=Ri$ across the load resistor $R$, the noise voltage at the detector output is
\begin{align}
S_v=2qR\overline{v}\qquad\unit{V^2/Hz}.
\end{align}
Then, the shot noise is referred to the input-power noise using $v=k_dP$.  Thus, at the operating power $P_0$ it holds that
\begin{align}
S_P=2q\frac{RP_0}{k_d}\qquad\text{shot noise,}~\unit{W^2/Hz}.
\label{eqn:am-shot-at-input}
\end{align}

The thermal noise across load resistance $R$ has the power spectral density
\begin{align}
S_v=4kTR\qquad\unit{V^2/Hz},
\label{eqn:am-thermal-noise}
\end{align}
which turns into
\begin{align}
S_P=\frac{4kTR}{k_d^2}\qquad\text{thermal noise,}~\unit{W^2/Hz}.
\label{eqn:am-thermal-at-input}
\end{align}
referred to the detector input.
An additional thermal-noise contribution comes from the dissipative resistance of the diodes.  This can be accounted for by increasing the value of $R$ in Equations~\req{eqn:am-thermal-noise} and \req{eqn:am-thermal-at-input}.   It should be remarked that diode differential resistance is not a dissipative phenomenon, for there is no thermal noise associated to it.

\begin{figure}[t]
  \centering\namedgraphics{0.8}{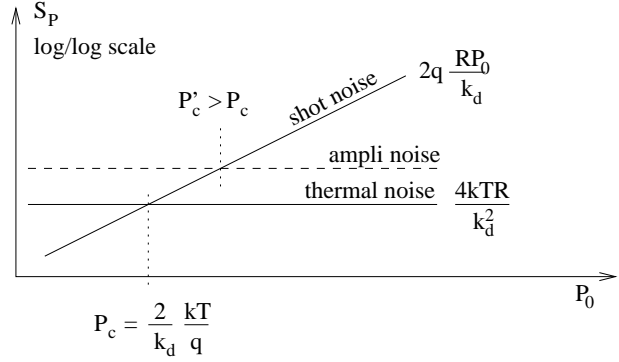}
\caption{Power spectral density of the detector noise, referred at the input.}
\label{fig:am-fundam-noise}
\end{figure}
Figure \ref{fig:am-fundam-noise} shows the equivalent input noise as a function of
power.  The shot noise is equal to the thermal noise, 
$(S_P)_\text{shot}=(S_P)_\text{thermal}$, at the critical power 
\begin{align}
P_c=\frac{2}{k_d}\,\frac{kT}{q}~.
\label{eqn:am-critical-p}
\end{align}

It turns out that the power detector is always used in the low power region ($P_0<P_c$), where shot noise is negligible.
In fact, taking the data of Table~\ref{tab:am-conversion-gain} as typical of actual detectors, $P_c$ spans from $+2$ dBm to $-10$ dBm for the Schottky diodes ($35\,\unit{A^{-1}}<k_d<500\,\unit{A^{-1}}$), and from $-8$ dBm to $-12$ dBm for the tunnel diodes ($330\,\unit{A^{-1}}<k_d<820\,\unit{A^{-1}}$), depending on the load resistance.  On the other hand, the detector turns from the quadratic (power) response to the linear (voltage) response at a significantly lower power.  This can be seen on Figure \ref{fig:am-dzr124aa} and \ref{fig:am-dt8012}.

Looking at the specifications of commercial power detectors, information about noise is scarce.
Some manufacturers give the NEP (Noise Equivalent Power) parameter, i.e., the power at the detector input that produces a video output equal to that of the device noise.  In no case is said whether the NEP increases or not in the presence of a strong input signal, which is related to precision.
Even worse, no data about flickering is found in the literature or in the data sheets.  Only one manufacturer (Herotek) claims the low flicker feature of its tunnel diodes, yet without providing any data.

The power detector is always connected to some kind of amplifier, which is noisy.
Denoting with $(h_0)_\text{ampli}$ and $(h_{-1})_\text{ampli}$ the  white and flicker noise coefficients of the amplifier, the spectrum density referred at the input is
\begin{align}
\label{eqn:am-ampli-noise}
S_P(f)&=\frac{(h_0)_\text{ampli}}{k^2_d} + \frac{(h_{-1})_\text{ampli}}{k^d_d}\,\frac1f~.
\end{align}
The amplifier noise coefficient $(h_0)_\text{ampli}$ is connected to the noise figure by
$(h_0)_\text{ampli}=(F-1)kT$.  Yet we prefer not to use the noise figure because in general the amplifier noise results from voltage noise and current noise, which depends on $R$. 
Equation ~\req{eqn:am-ampli-noise} is rewritten in terms of amplitude noise using $\alpha=\frac12\frac{\delta P}{P_0}$ [Eq.~\req{eqn:am-p-vs-alpha}].  Thus,
\begin{align}
\label{eqn:am-ampli-salpha}
S_\alpha(f)&= \frac{1}{2P_0}\,\frac{qR}{k_d} +
		\frac{1}{P_0^2}\,\frac{kTR}{k_d^2} + 
		\frac{1}{4P_0^2}\,\frac{(h_0)_\text{ampli}}{k^2_d} +
		\frac{1}{4P_0^2}\,\frac{(h_{-1})_\text{ampli}}{k^2_d}\,\frac1f~.
\end{align}

After the first term of Eq.~\req{eqn:am-ampli-noise}, the critical power becomes
\begin{align}
P'_c&=\frac{2}{k_d}\,\frac{kT}{q} + \frac{(h_0)_\text{ampli}}{2qRk_d}~.
\end{align}
This reinforces the conclusion that in actual conditions the shot noise is negligible.

\section{Design of the front-end amplifier}\label{sec:am-amplifier}
\begin{figure}[t]
\centering\namedgraphics{0.80}{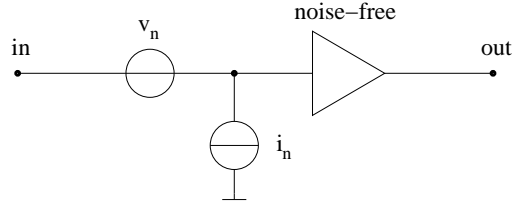}
\caption{Rothe-Dahlke model of the noisy amplifier.}
\label{fig:am-rothe-dahlke}
\end{figure}

For optimum design, one should account for the detector noise and for the noise of the amplifier, and find the most appropriate amplifier and operating conditions.  Yet, the optimum design relies upon the detailed knowledge of the power-detector noise, which is one of our targets (Sec.~\ref{sec:am-detector-noise-meas}).  
Thus, we provisionally neglect the excess noise of the power detector.  The first design is based on the available data, i.e., thermal noise and the noise of the amplifier.  
Operational amplifiers or other types of impedance-mismatched amplifiers are often used in practice.  As a consequence, a single parameter, i.e., the noise figure or the noise temperature, is not sufficient to describe the amplifier noise.  Voltage and current fluctuations must be treated separately, according to the popular Rothe-Dahlke model \cite{rothe56ire} (Fig.~\ref{fig:am-rothe-dahlke}).
The amplifier noise contains white and flicker, thus
\begin{align}
(S_v)_\text{ampli}&=h_{0,v}+h_{-1,v}\:\frac1f\\
(S_i)_\text{ampli}&=h_{0,i}+h_{-1,i}\:\frac1f~.
\end{align}
The design can be corrected afterwards, accounting for the flicker noise of the detector.

\subsection*{Single-channel systems}
Accounting for shot and thermal noise, and for the noise of the amplifier, the noise spectrum density is 
\begin{align}
S_v&=2qR\overline{v} + 4kTR + 
\left(S_v\right)_\text{ampli} + R^2\left(S_i\right)_\text{ampli}
\end{align}
at the amplifier input, and
\begin{align}
S_P&=2q\frac{RP}{k_d} + \frac{4kTR}{k^2_d} + 
\frac{\left(S_v\right)_\text{ampli}}{k^2_d} + 
\frac{R^2\left(S_i\right)_\text{ampli}}{k^2_d}
\end{align}
referred to the rf input.  The detector gain $k_d$ depends on $R$, thus the residual $S_P$ can not be arbitrarily reduced by decreasing $R$.  Instead, there is an optimum $R$ at which the system noise is at its minimum.

\subsection*{Correlation-and-averaging systems}
\begin{figure}[t]
\centering\namedgraphics{0.80}{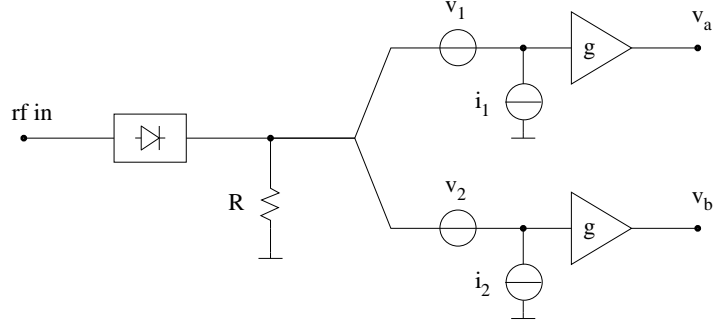}
\caption{The load resistor turns the current noise into fully-correlated noise.}
\label{fig:am-correl-noise}
\end{figure}
The noise contribution of the amplifier can be reduced by measuring the cross spectrum at the output of two amplifiers connected to the power detector, provided that the noise of the amplifiers is independent.  For this to be true, the optimum design of the front-end amplifier changes radically.  Figure~\ref{fig:am-correl-noise} points out the problem.  
The current noise of each amplifier turns into a random voltage fluctuation across the load resistance $R$.   Focusing only on the amplifier noise, the voltage at the two outputs is
\begin{align*}
v_a&=g\left(v_1+Ri_1+Ri_2\right)\\
v_b&=g\left(v_2+Ri_1+Ri_2\right)
\end{align*}
The terms $gv_1$ and $gv_2$ are independent, for their contribution to the cross spectrum density is reduced by a factor $\smash{\frac{1}{\sqrt{2m}}}$, where $m$ is the number of averaged spectra.  Conversely, a term
\begin{align*}
gR(i_1+i_2)
\end{align*}
is present at the two outputs.  This term is exactly the same, thus it can not be reduced by correlation and averaging.
Consequently, the lowest current noise is the most important parameter, even if this is obtained at expense of a larger voltage noise.  Yet, the rejection of larger voltage noise requires large $m$, for some tradeoff may be necessary.

\subsection*{Examples}
This section shows some design attempts, aimed at the lowest white and flicker noise at low Fourier frequencies, up to 0.1--1 MHz, where operational amplifiers can be exploited in a simple way.  

A preliminary analysis reveals that, at the low resistance values required by the detector, BJT amplifiers perform lower noise than field-effect transistors.  On the other hand, the noise rejection by correlation and average requires low current noise, for JFET amplifiers are the best choice.  In fact, BJTs can not be used because of the current noise, while MOSFETs show $1/f$ noise significantly larger (10 dB or more) than JFETs.  

\begin{table}
\caption{Noise parameters of some selected amplifiers.}
\label{tab:am-ampli-noise}
\def\rrc{\rule[-1.0ex]{0ex}{3.3ex}}
\vspace*{0.5ex}\centering%
\begin{tabular}{|c|c|c|c|c|l|}\hline
\rrc			& \multicolumn{2}{c|}{voltage} & \multicolumn{2}{c|}{current}&\\\cline{2-5}  
\rrc type	& white & flicker & white & flicker & notes \\ 
\rrc			& $h_{0,v}$ & $h_{-1,v}$ & $h_{0,i}$ & $h_{-1,i}$ &\\\hline
\rrc AD743	& 2.9 & 18  & 6.9    & $-$ & jfet op-amp\\\hline
\rrc LT1028	& 0.9 & 1.7 & 1000 & 16   & bjt op-amp\\\hline
\rrc MAT02	& 0.9 & 1.6 & 900   & 1.6  & npn bjt matched pair\\\hline
\rrc MAT03	& 0.7 & 1.2 & 1200 & 11   & pnp bjt matched pair\\\hline
\rrc OP27	& 3.0 & 4.3 & 400   & 4.7  & bjt op-amp\\\hline
\rrc OP177	& 10  & 8.0 & 125   & 1.6  & bjt op-amp\\\hline
\rrc OP1177& 8.0 & 8.3 & 200   & 1.5  & bjt op-amp\\\hline
\rrc OPA627& 4.5 & 45  & 2.5    & $-$ & jfet op-amp\\\hline
\rule[-2.0ex]{0ex}{4.5ex}
unit	&\unit{\frac{nV}{\sqrt{Hz}}} &\unit{\frac{nV}{\sqrt{Hz}}} &
		\unit{\frac{fA}{\sqrt{Hz}}}  & \unit{\frac{pA}{\sqrt{Hz}}} & \\\hline
\end{tabular}
\end{table}

Using two detectors (DZR124AA and DT8012), we try the operational amplifiers and transistor pairs listed on Table~\ref{tab:am-ampli-noise}.  These amplifiers are selected with the criterion that each one is a low-noise choice in its category.
\begin{description}
\item[AD743 and OPA627] are general-purpose precision JFET amplifiers, which exhibit low bias current, hence low current noise.  They are intended for correlation-and-averaging schemes (Fig.~\ref{fig:am-correl-noise}).  The \textbf{OP625} is similar to the OP627 but for the frequency compensation, which enables unity-gain operations, yet at expenses of speed.  It is used successfully in the measurement of the excess noised of semiconductors, where large averaging size is necessary in order to rid of the amplifier noise \cite{sampietro99rsi}.
\item[LT1028] is a fast BJT amplifier with high bias current in the differential input stage.  This feature makes it suitable to low-noise applications in which the source resistance is low.  In fact, the optimum noise resistance $R_b=\sqrt{h_v/h_i}$ is of 900 \ohm\ for white noise, and of 105 \ohm\ for flicker.  These values are in the preferred range for proper operation of the power detectors.    
\item[MAT02 and MAT03] are bipolar matched pairs.  They exhibit lower noise than operational amplifiers, and they are suitable to the design for low resistance of the source, like the LT1028.  The MAT03 was successfully employed in the design of a low-noise amplifier optimized for 50 \ohm\ sources \cite{rubiola04rsi-amplifier}.
\item[OP27 and OP37] are popular general-purpose precision BJT amplifiers, most often used in low-noise applications.  Their noise characteristics are about identical. The OP27 is fully compensated, for it is stable at closed-loop gain of one.  The OP37 is only partially compensated, which requires a minimum closed-loop gain of five for stable operation.  Of course, lower compensation increases bandwidth and speed.
\item[OP177 and OP1177] are general-purpose precision BJT amplifiers with low bias current in the differential input stage, thus they exhibit lower current noise than other BJT amplifiers.  They can be an alternative if the design based on JFET amplifier fails.
\end{description}

\noindent
In the try-and-error process, we take into account shot noise, thermal noise, amplifier voltage noise, and amplifier current noise.
The total white noise is the sum of all them
\begin{align}
S_P&=2q\frac{RP}{k_d} + \frac{4kTR}{k^2_d} + 
	\frac{h_{0,v}}{k^2_d} + \frac{R^2\,h_{0,i}}{k^2_d}
	&&\text{total white [\,$+$\,]}~.
	\label{eqn:am-design-total-white}
\intertext{The total flicker accounts for the voltage and current of the amplifier}
S_P&=\frac{h_{-1,v}}{k^2_d} + \frac{R^2\,h_{-1,i}}{k^2_d}
	&&\text{total flicker [\,$\Diamond$\,]}~.
	\label{eqn:am-design-total-flicker}
\intertext{The correlated noise spectrum is equal to the total noise minus the voltage noise spectrum of the amplifier, which is independent}
S_P&=2q\frac{RP}{k_d} + \frac{4kTR}{k^2_d} +
	\frac{R^2\,h_{0,i}}{k^2_d}&&\text{correlated white [\,{\footnotesize$\bigcirc$}\,]}~,
		\label{eqn:am-design-correl-white}\\
S_P&=\frac{R^2\,h_{-1,i}}{k^2_d}&&\text{correlated flicker [\,$\Box$\,]}~.
		\label{eqn:am-design-correl-flicker}
\end{align}
The flicker noise of the detector, still not available, is to be added to Equations \req{eqn:am-design-total-flicker} and \req{eqn:am-design-correl-flicker}.

\def\myplot#1{\includegraphics[scale=0.385]{#1}}
\def\legenda{%
\begin{tabular}{llcll}
$+$ &white noise&&$\Diamond$& flicker noise\\ 
{\footnotesize$\bigcirc$}&correlated white noise &&$\Box$& correlated flicker noise ($i$ only)
\end{tabular}}

\begin{figure}[t]
\vfill\centering%
\myplot{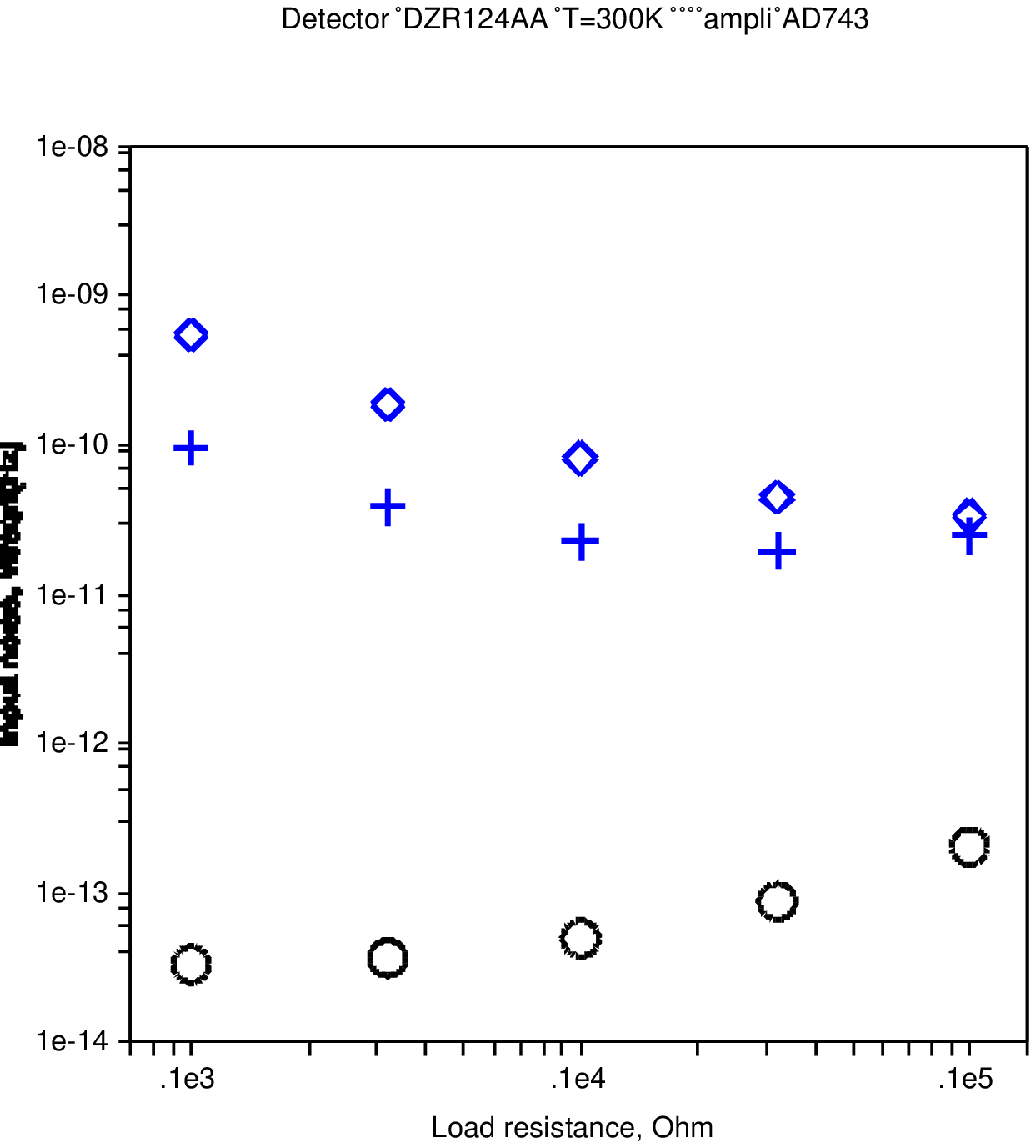}~~\myplot{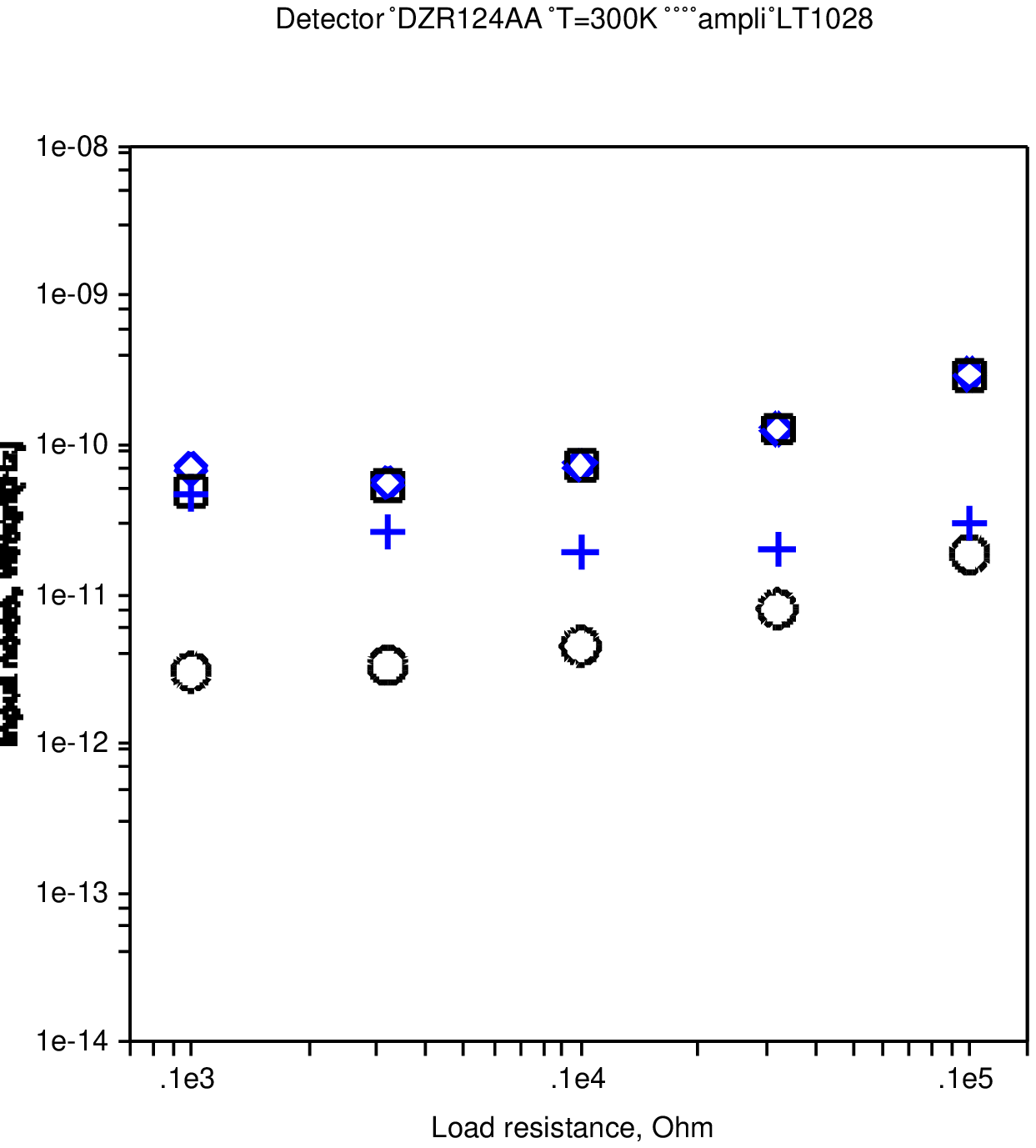}\\[1ex]
\myplot{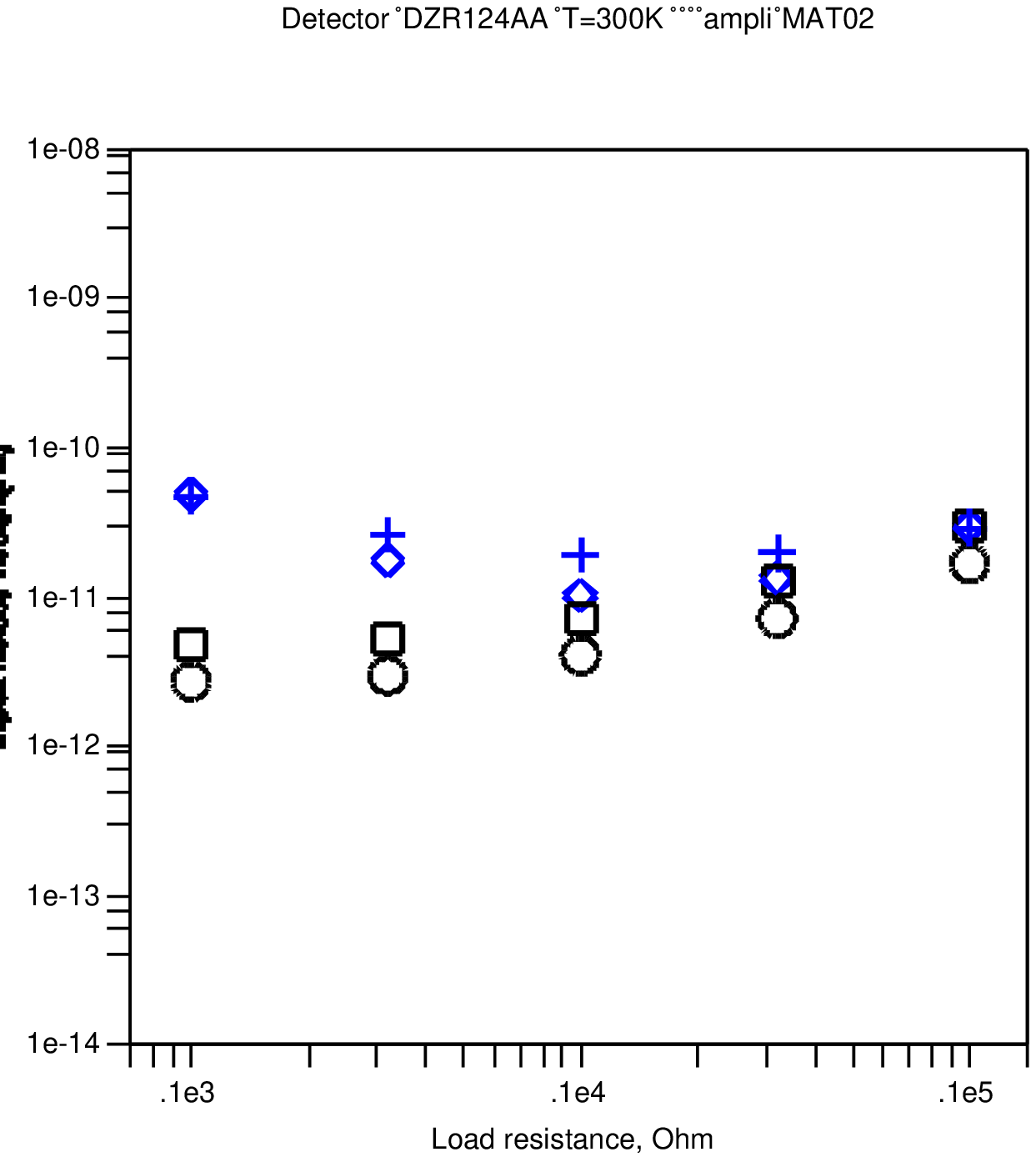}~~\myplot{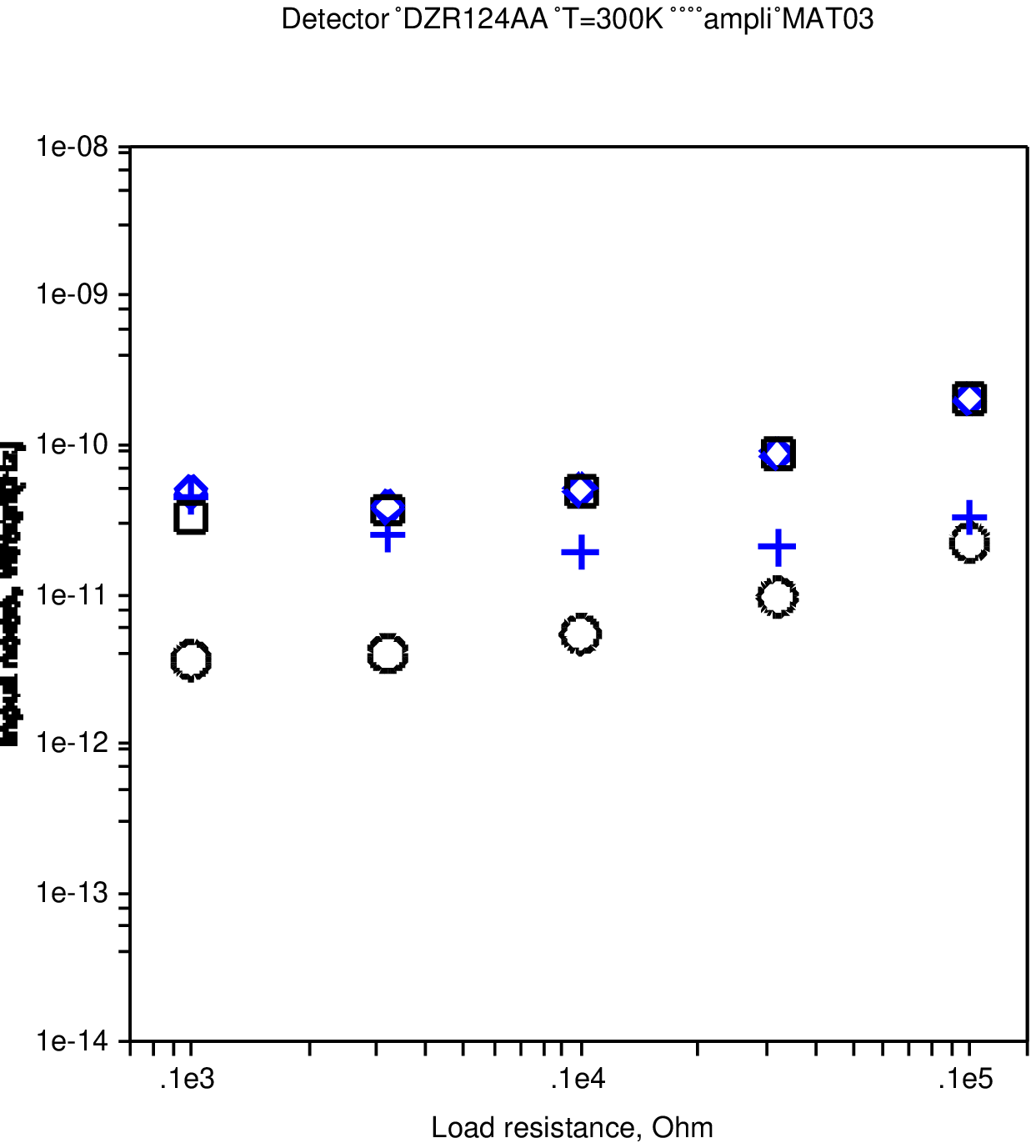}\\[1ex]
\myplot{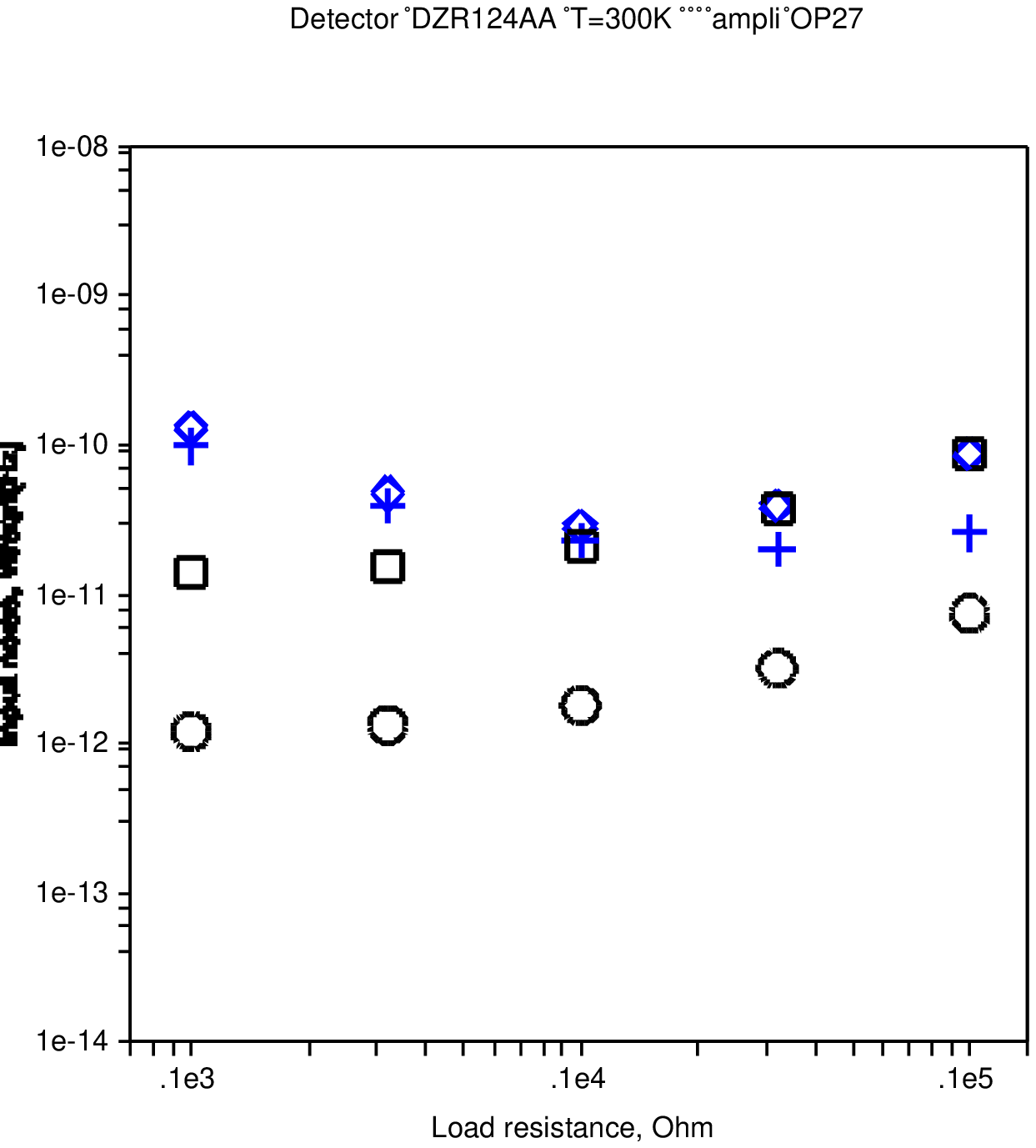}~~\myplot{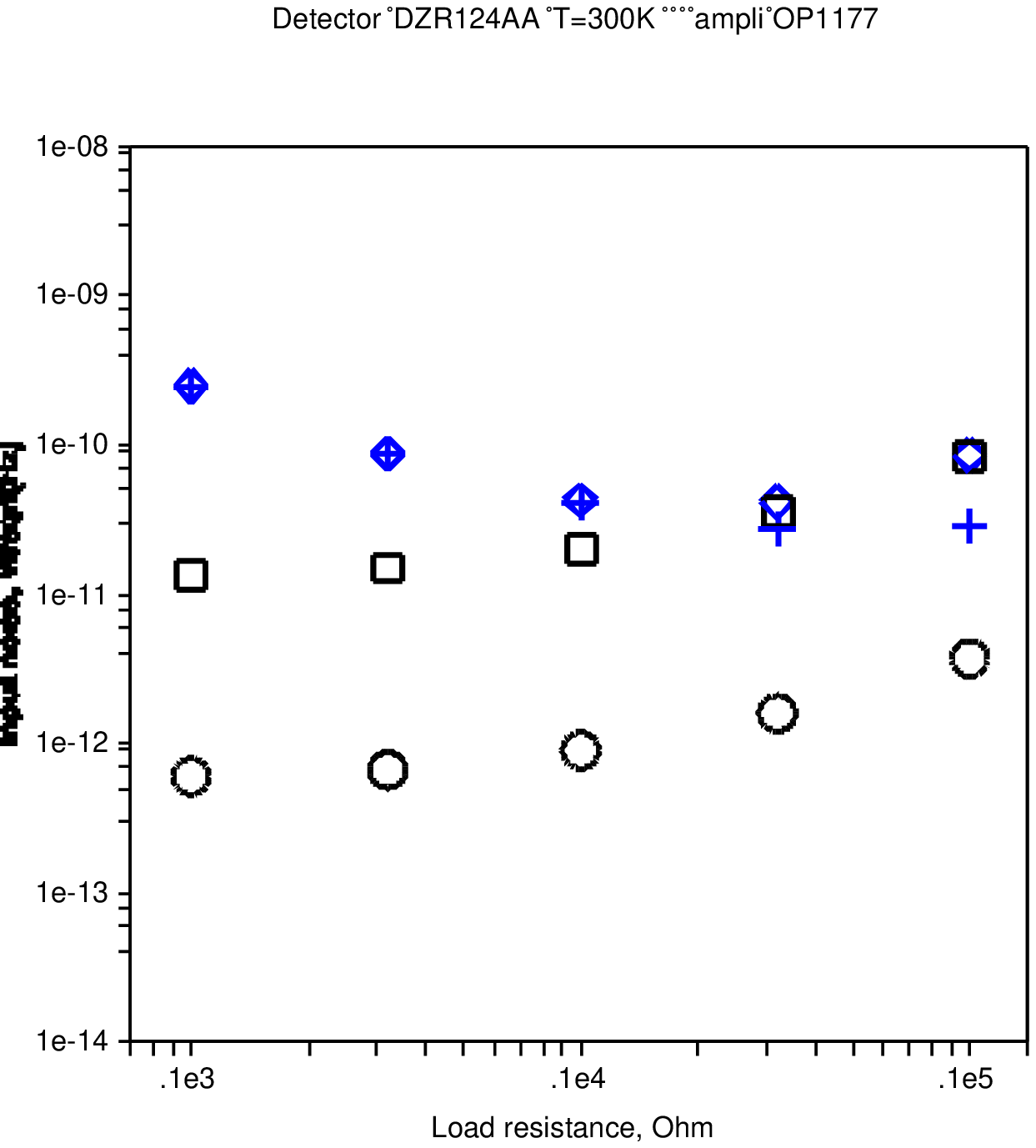}\\
\legenda
\caption{Expected noise of the DZR124AA (Schottky) followed an amplifier.}
\label{fig:am-dzr124aa-noise}
\end{figure}

\begin{figure}[t]
\vfill\centering%
\myplot{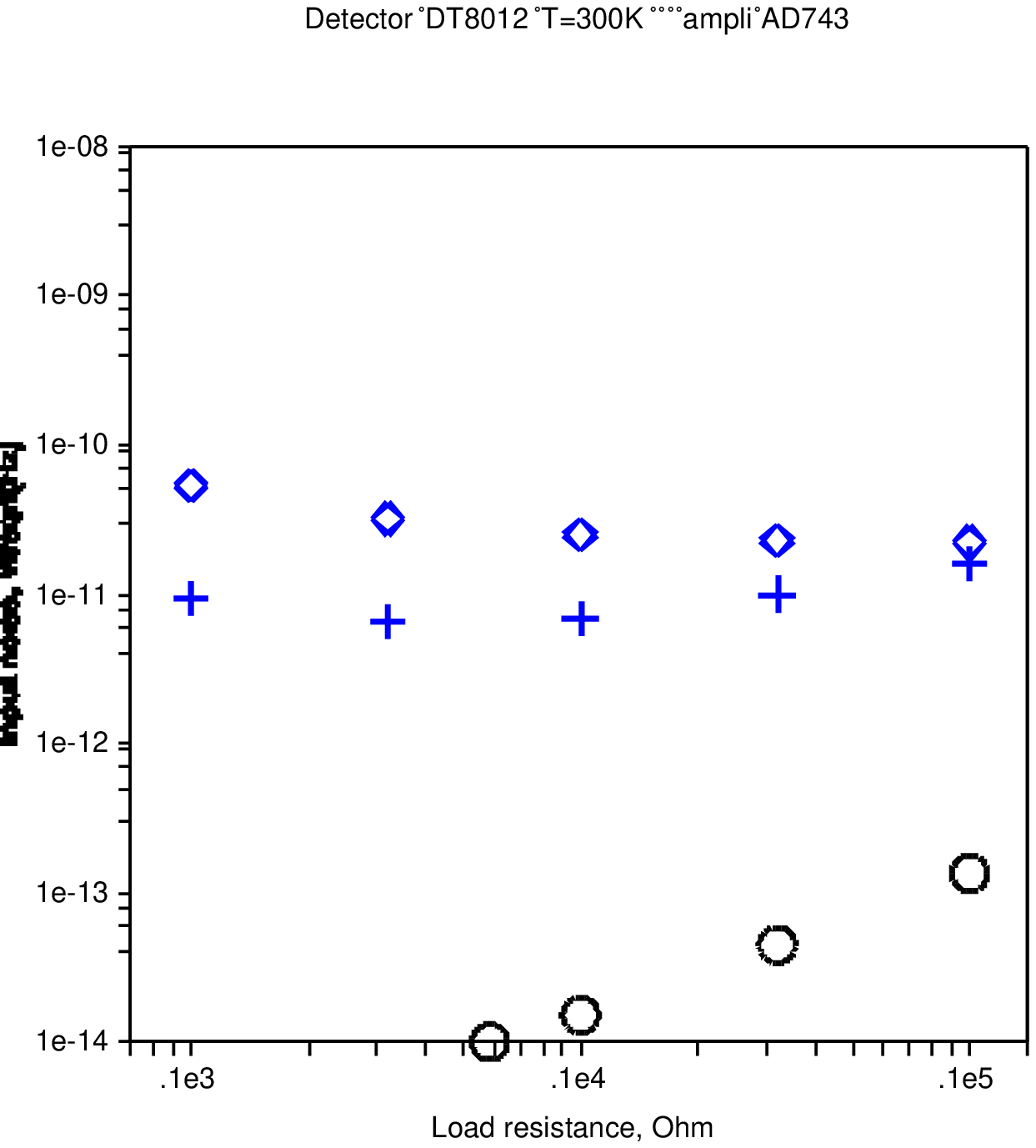}~~\myplot{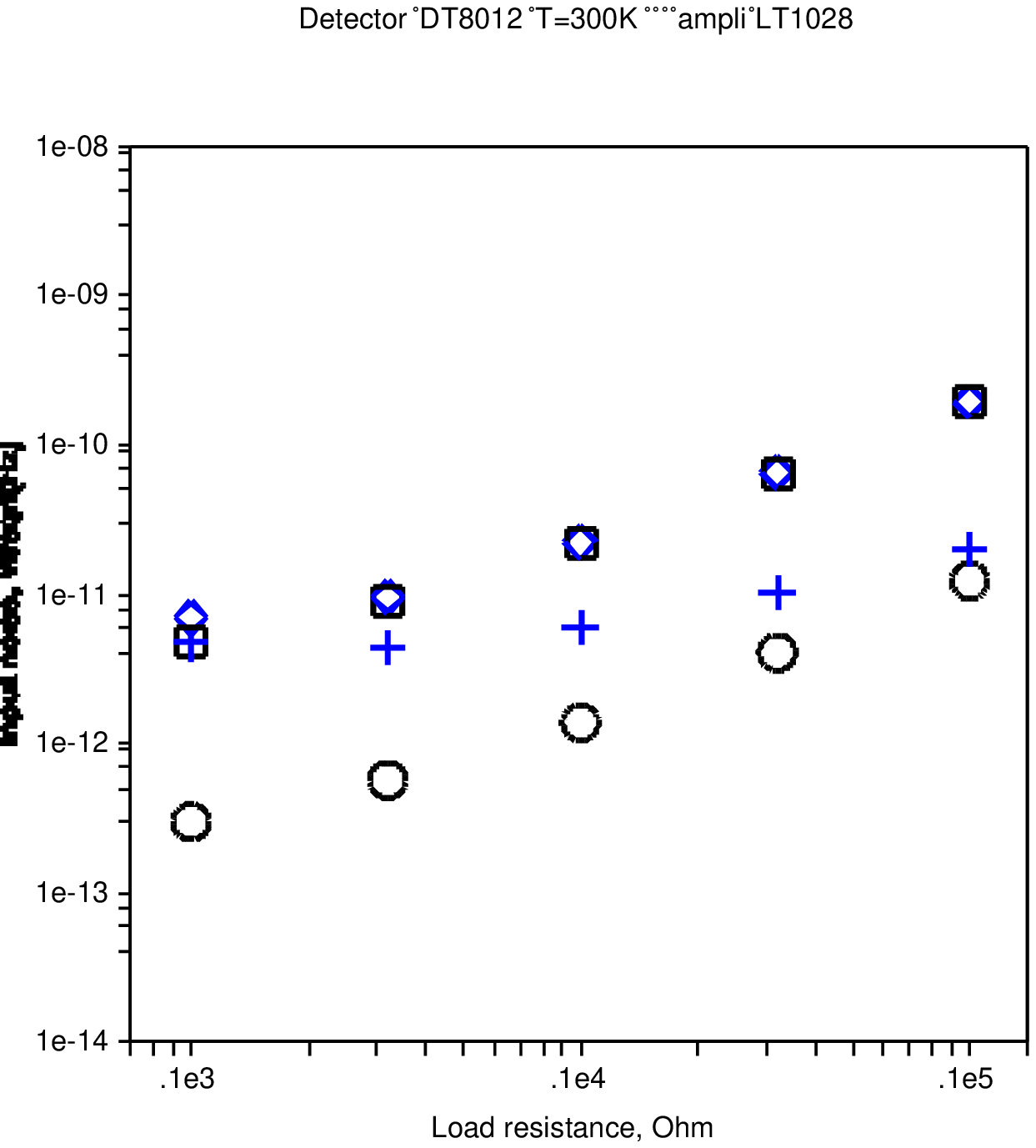}\\[1ex]
\myplot{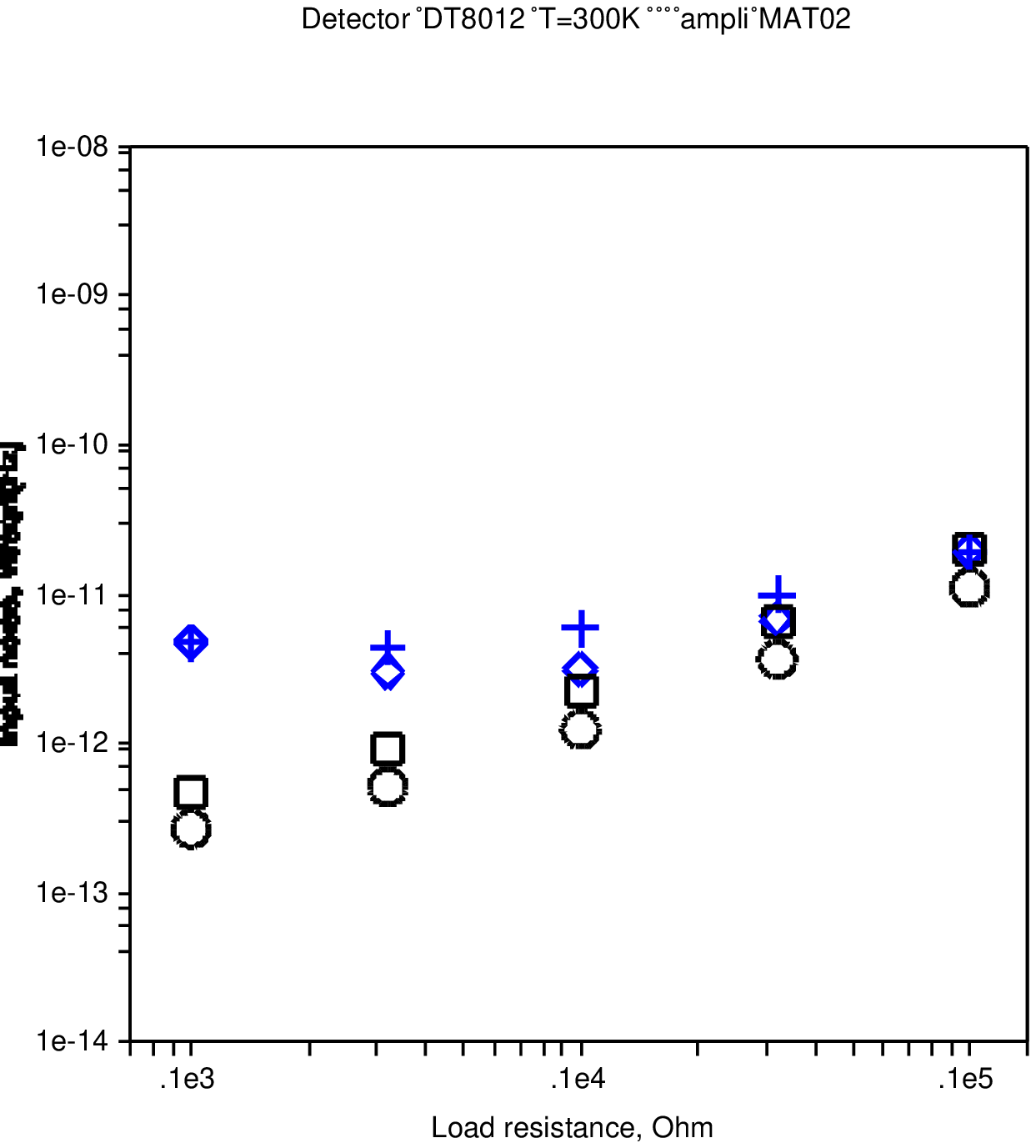}~~\myplot{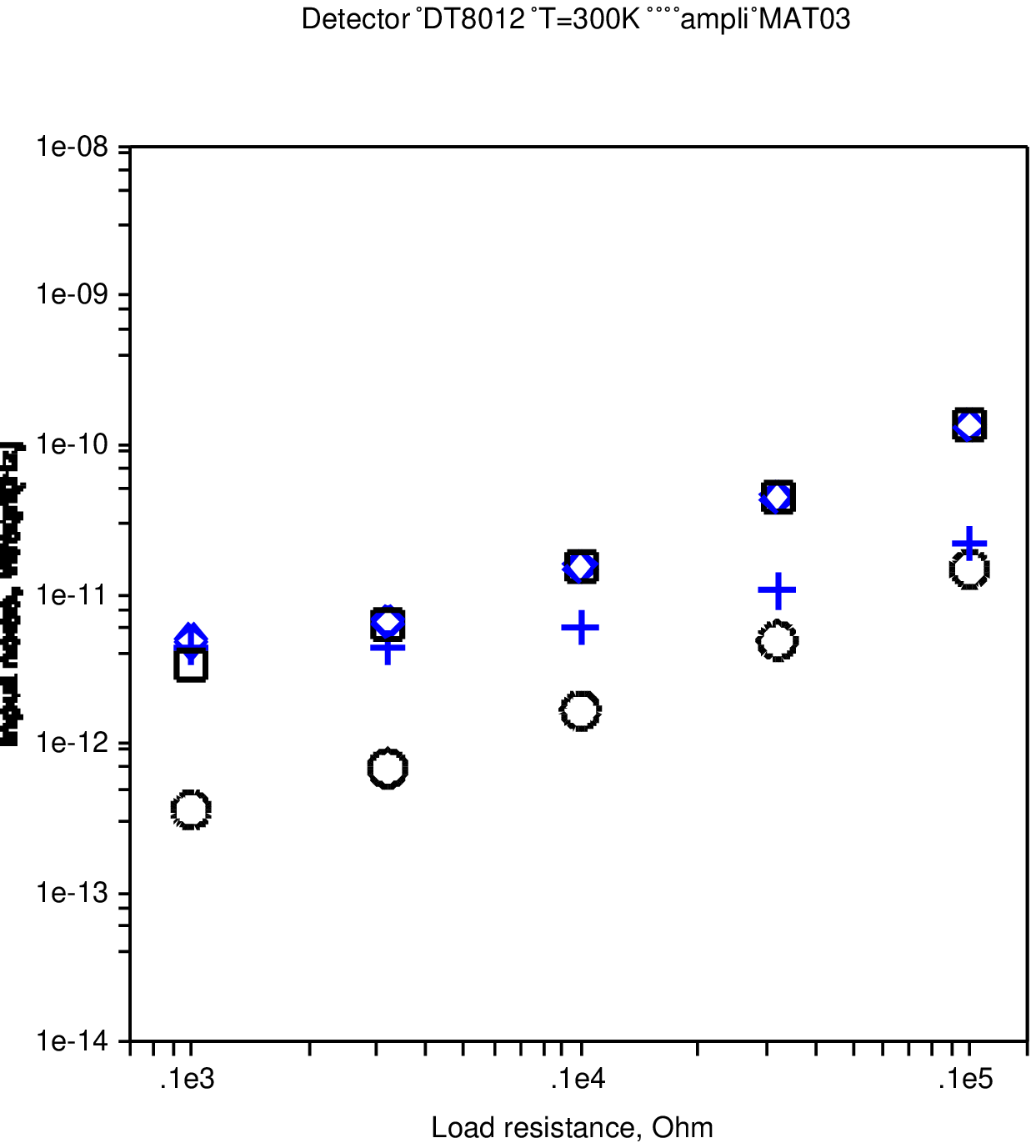}\\[1ex]
\myplot{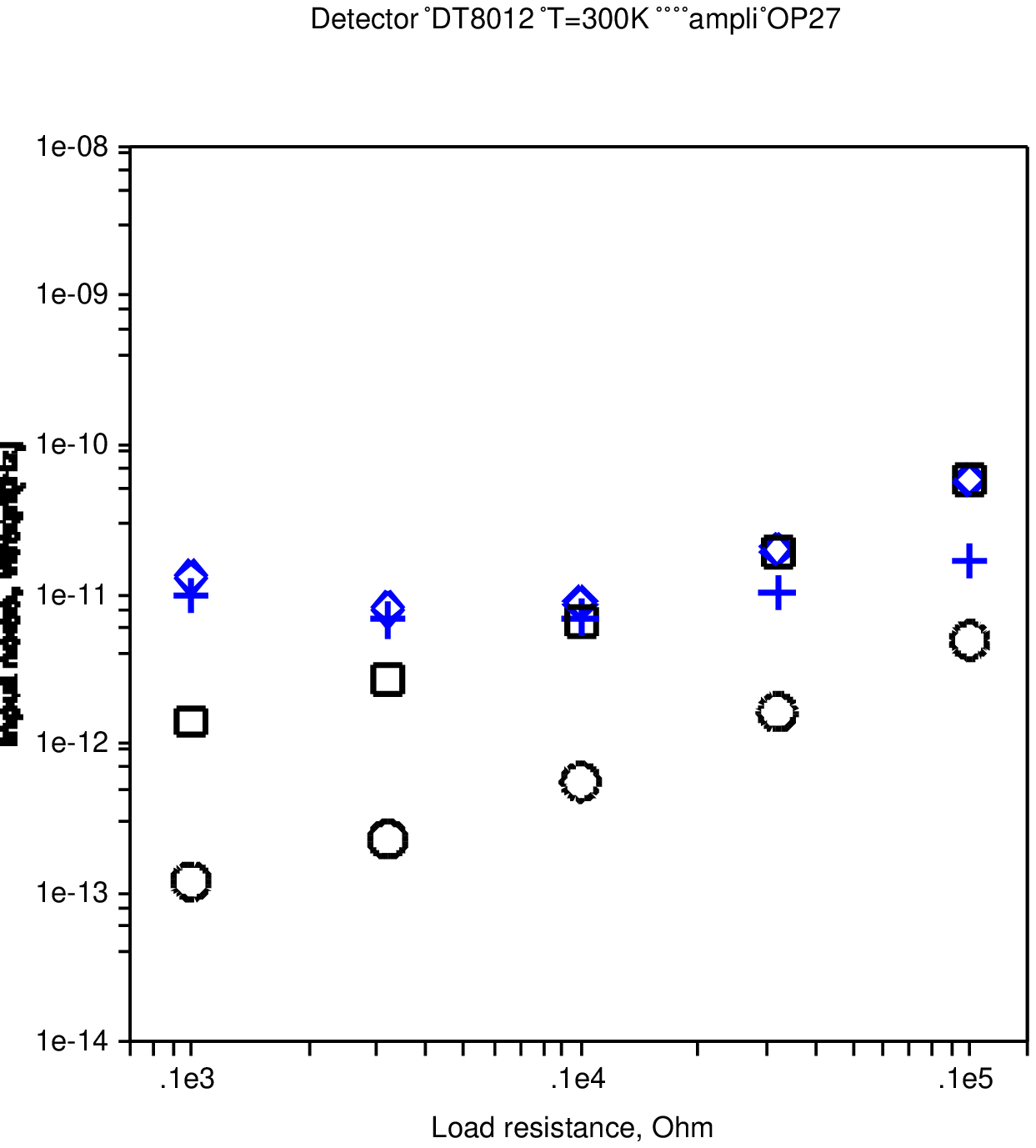}~~\myplot{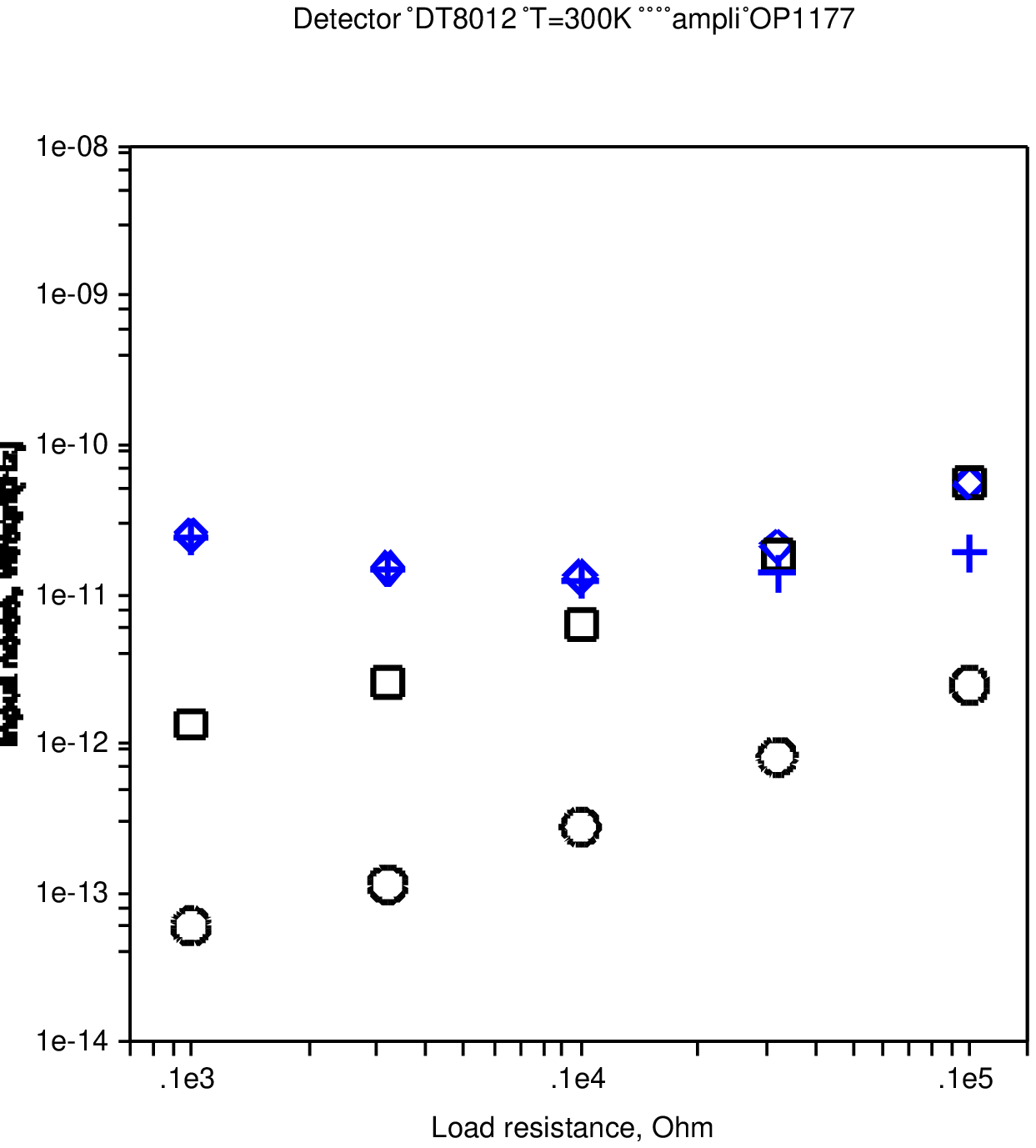}\\
\legenda
\caption{Expected noise of the DT8012 (Tunnel) followed an amplifier.}
\label{fig:am-dt8012-noise}
\end{figure}

After evaluation, we discard the OP177 because the OP1177 exhibits superior performances in all noise parameters.  Similarly, we discard the OP627 in favour of the AD743.   The lower current noise of the OP627 can be exploited only at very large number of averages, which is impractical.  On the other hand, the lower voltage noise of the AD743 helps to keep the number of averages reasonable.
Figures \ref{fig:am-dzr124aa-noise} and \ref{fig:am-dt8012-noise} show a summary of the expected noise as a function of the load resistance $R$.
The symbols $+$, {\footnotesize$\bigcirc$}, $\Diamond$, $\Box$ have the same meaning in Equations \req{eqn:am-design-total-white}--\req{eqn:am-design-correl-flicker} and in Figures \ref{fig:am-dzr124aa-noise} and \ref{fig:am-dt8012-noise}.  Analyzing the noise plots we restrict our attention to the detector-amplifier pairs of Table \ref{tab:am-expected-noise}, in which we identify the following interesting configurations. 

\ifx\undefined\unit\def\unit#1{\ensuremath{\mathrm{\,#1}}}\fi
\ifx\undefined\ohm\def\ohm{\ensuremath{\mathrm{\Omega}}}\fi
\definecolor{light}{gray}{0.80}
\def\cb#1{\smash{\colorbox{light}{\parbox{5ex}{\centering#1}}}}
\def\rrt{\rule[0ex]{0ex}{2.5ex}}
\def\rrc{\rule[-1.5ex]{0ex}{4ex}}
\def\rrb{\rule[-1.5ex]{0ex}{1.5ex}}
\def\rrr{\rule[0ex]{0ex}{1.5ex}}
\def\devs#1#2{\smash{\raisebox{-4ex}{\hspace*{-1.5ex}\begin{tabular}{c}#1\\[-0.3ex]and\\#2\end{tabular}\hspace*{-1.5ex}}}}
\def\sunit{\smash{\raisebox{-4ex}{\hspace*{-1.5ex}\begin{tabular}{c}$10^{-12}$\\[0.5ex]\unit{\dfrac{W}{\sqrt{Hz}}}\end{tabular}\hspace*{-1.5ex}}}}
\begin{table}[t]
\caption{Expected noise of a power detector followed by an amplifier.  
The grey box shows the best choice for the amplifier-detector pair.}
\label{tab:am-expected-noise}
\vspace*{0.5ex}\small\centering%
\begin{tabular}{|c|c|ccccc|c|}\hline\hline
\multicolumn{8}{|c|}{{\rrc Equivalent input noise}}\\\hline\hline
\smash{\raisebox{-2ex}{\hspace*{-1.5ex}\begin{tabular}{c}detector\\[-0.5ex]and\\[-0.2ex]amplifier\end{tabular}\hspace*{-1.5ex}}}
&\smash{\raisebox{-2ex}{noise type}}
		& \multicolumn{5}{c|}{load resistance} &{\rrc}unit\\\cline{3-8}
&		&100   & 320  &1000 &3200 &10000&{\rrc}\ohm\\\hline\hline
\devs{DZR124AA}{AD743}
&white{\rrt}		& 99.2 & 39.5 & 22.8 &\cb{19.6}&  24.5 &\sunit\\
&flicker		& 541  & 186  & 80.8 &\cb{44.8}& 33.3  &\\
&correl.\ white	& 0.050&0.042&0.052&\cb{0.089}&0.205 &\\
&correl.\ flicker\rrb	& --      & --     & --     &\cb{ --\rrr }   & --     &\\
\hline\hline
\devs{DZR124AA}{MAT02}
&white{\rrt}	& 54.7 & 27.5 &\cb{19.6}&19.8  &  29.2 &\sunit\\
&flicker			& 48.6 & 17.7 &\cb{10.4}&13.5  &  29.9 &\\
&correl.\ white	& 4.06 &  3.45&  4.24&7.28  &  16.7 &\\
&correl.\ flicker\rrb&7.22&6.13&  7.54& 12.9 &  29.8 &\\
\hline\hline
\devs{DZR124AA}{OP27}
&white{\rrt}		& 102  & 40.3 &\cb{23.1}& 20.0 & 25.6  &\sunit\\
&flicker		& 131  & 48.0 &\cb{29.4}& 39.5 & 87.8  &\\
&correl.\ white	& 1.80 & 1.53 &\cb{1.88} &  3.23& 7.44  &\\
&correl.\ flicker\rrb	&21.2	 &18.0  &\cb{22.1}& 38.0 & 87.4  &\\
\hline\hline
\devs{DT8012}{AD743}
&white{\rrt}	& 9.86 &  6.77&\cb{7.00}& 10.0 & 16.2  &\sunit\\
&flicker			& 53.7 & 32.0 &\cb{24.8}& 22.8 & 22.1  &\\
&correl.\ white	&0.005&0.007 &0.016&0.045& 0.135&\\
&correl.\ flicker\rrb	& --      & --     & --     & --     & --     &\\
\hline\hline
\devs{DT8012}{LT1028}
&white{\rrt}	&\cb{5.44}& 4.72 & 6.06 & 10.2 & 20.1  &\sunit\\
&flicker			&\cb{8.78}& 10.9 & 23.3 & 66.0 & 197   &\\
&correl.\ white	&0.448& 0.657& 1.45 & 4.12 & 12.3  &\\
&correl.\ flicker\rrb&7.17&10.5& 23.2 & 66.0 & 197   &\\
\hline\hline
\devs{DT8012}{MAT02}
&white{\rrt}		&\cb{5.43}&\cb{4.72}&  6.02& 10.1 & 19.3  &\sunit\\
&flicker		&\cb{4.83}&\cb{3.03}&  3.20&  6.90&  19.8 &\\
&correl.\ white	&\cb{0.403}& 0.591&  1.30& 3.71 & 11.1  &\\
&correl.\ flicker\rrb&\cb{0.717}&1.05& 2.32&6.60 & 19.7  &\\
\hline\hline
\multicolumn{8}{|c|}{\rrt%
White and shot noise, plus white and flicker noise of the amplifier.}\\  
\multicolumn{8}{|c|}{\rrb%
The flicker noise of the power detector is not accounted for.}\\
\hline\hline
\end{tabular}
\end{table}

\paragraph{Detector DZR124AA (Schottky)}\mbox{}
\begin{description}
\item[AD743 and 3.2 k\ohm.] Best design for a correlation-and-averaging system.
In principle, lower white noise can be achieved with lower load resistance, yet only with $m>10^4$.
\item[MAT02 and 1 k\ohm.] Lowest flicker in real-time measurements (single-channel). 
\item[OP27 and 1 k\ohm.] Simple nearly-optimum design if one is interested to white noise.  In fact, the white noise is 1.5 dB higher than that of the two above configurations;  this gap increases only at $m>300$ if the AD743 (jfet) is used.  On the other hand, the flicker noise is high and can not reduced by correlation.
\end{description}

\paragraph{Detector DT8012 (tunnel)}%
\begin{description}
\item[AD743 and 1 k\ohm.] Best design for a correlation-and-averaging system.
Slightly lower white noise can be obtained at lower $R$, yet at expenses of larger $m$ and of larger flicker noise.
\item[LT1028 and 100 \ohm.] Simple nearly-optimum design for real-time (single channel) systems.  This configuration, as compared to the best one (MAT02 with 320 \ohm\ load) shows white noise 1.2 dB higher, and flicker noise 9 dB higher. 
\item[MAT02 and 320 \ohm.] Lowest white and flicker noise in real-time measurements (single-channel). 
\item[MAT02 and 100 \ohm.]  Close to the lowest white and flicker noise in real-time measurements (single-channel).  Fairly good for correlation at moderate averaging, up to $m=360$ for white noise, and $m=90$ for flicker.
\end{description}

\paragraph{Remark.}  Generally, tunnel detectors show higher gain than Schottky detectors.  The fact that they exhibit lower noise is a consequence.  On the other hand, the Schottky detectors are often preferred because of wider bandwidth, and because of higher tolerance to electrical stress and to experimental errors.

\begin{figure}[t]
\centering\namedgraphics{0.70}{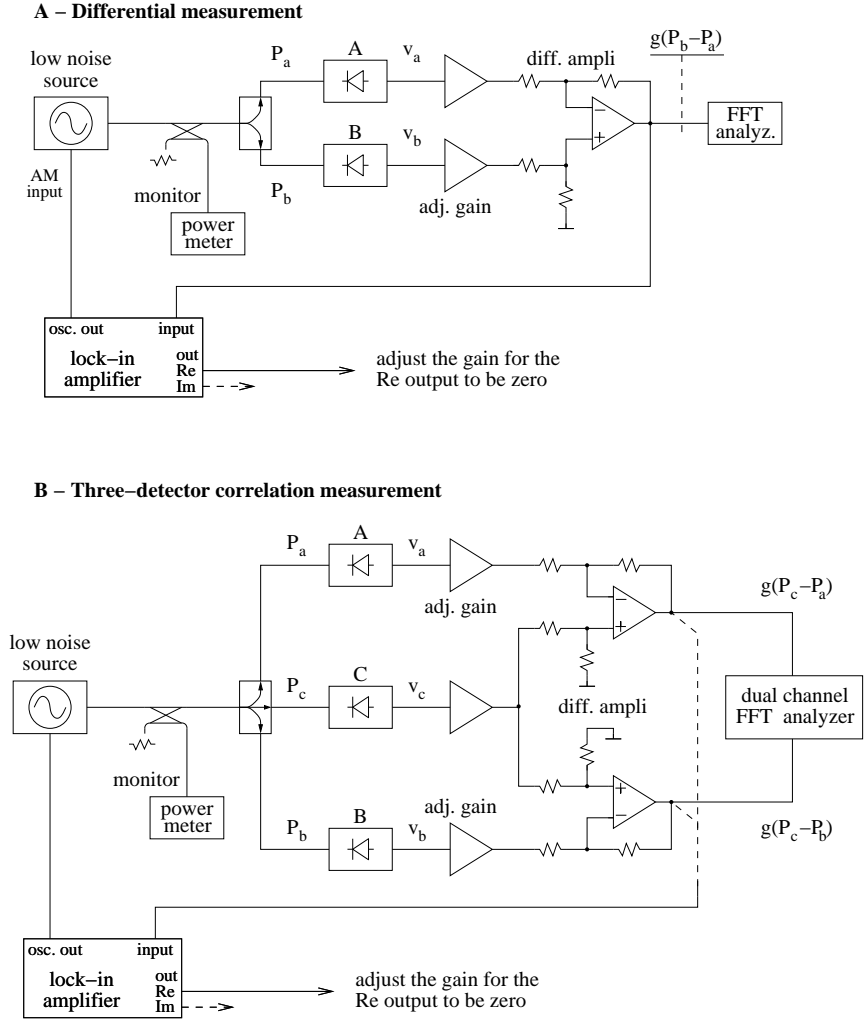}
\caption{Measurement of the power detector noise.}
\label{fig:am-detector-meas}
\end{figure}
\section{The measurement of the power detector noise}\label{sec:am-detector-noise-meas}
A detector alone can be measured only if a reference source is available whose AM noise is lower than the detector noise, and if the amplifier noise can be made negligible.  These are unrealistic requirements.  

It useful to compare two detectors, as in Fig.~\ref{fig:am-detector-meas}\,A.  The trick is to measure a differential signal $g(P_b-P_a)\approx0$, which is not affected by the power fluctuation of the source.  The lock-in helps in making the output independent of the power fluctuations.  Some residual PM noise has no effect on the detected voltage.  

One problem with the scheme of Fig.~\ref{fig:am-detector-meas}\,A
is that the measured noise is the sum of the noise of the two detectors, for the result relies upon the assumption that the two detectors are about equal.
This is fixed with the correlation scheme of Fig.~\ref{fig:am-detector-meas}\,B)\@.  The detector \emph{c}  is the device under test, while the two other detectors are used to cancel the fluctuations of the input power.
Thus
\begin{align*}
&g(P_c-P_a)  \qquad \text{and}\qquad g(P_c-P_b)
\end{align*}
After rejecting the single-channel noise by correlation and averaging, there results the noise of the detector C.

\begin{figure}[t]
\centering\namedgraphics{0.65}{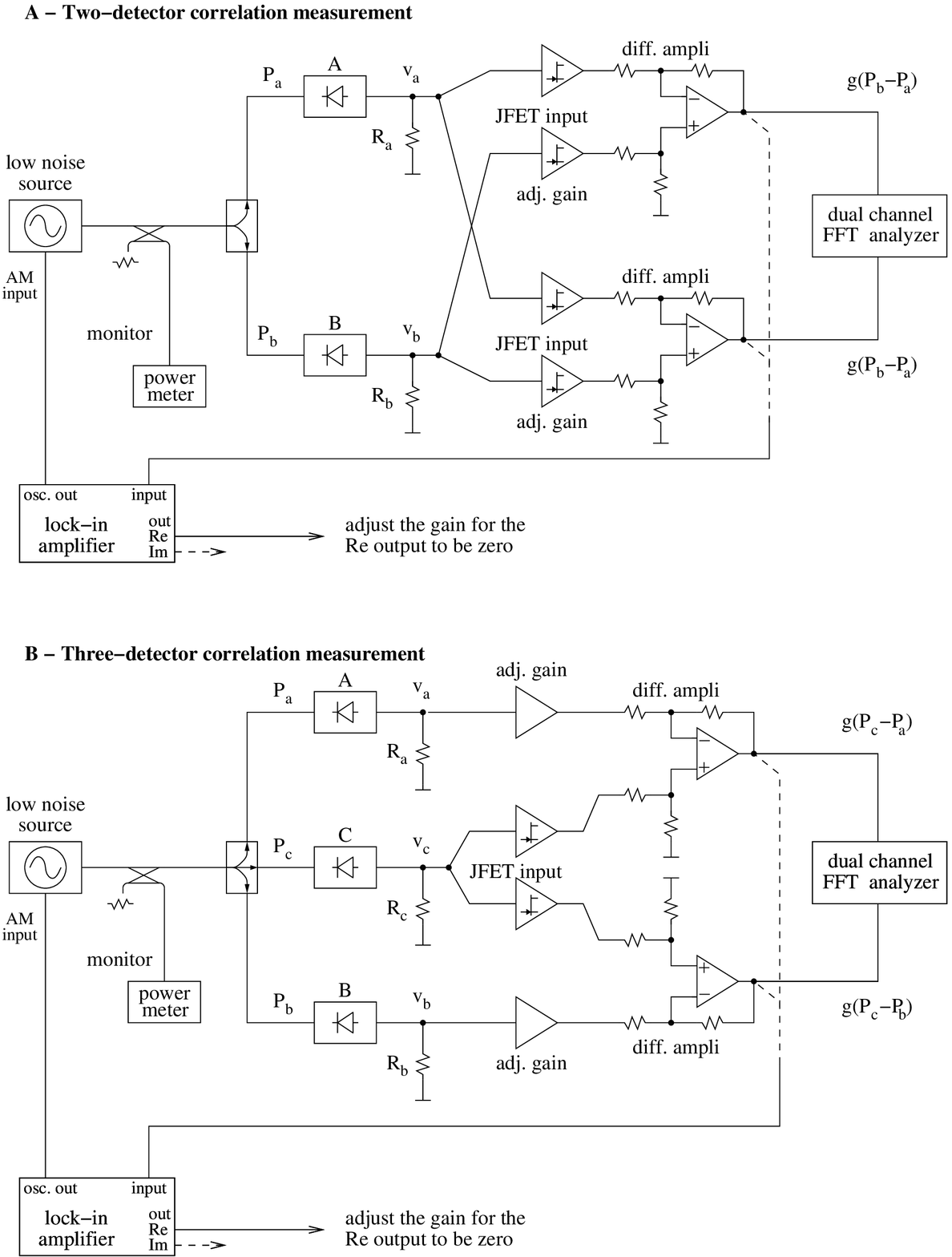}
\caption{Improved scheme for the measurement of the power detector noise.}
\label{fig:am-detector-meas-true}
\end{figure}
Another problem with the schemes of Fig.~\ref{fig:am-detector-meas} is that the noise of the amplifier is taken in.  This is fixed by using two independent amplifiers at the output of the power detector, as in Fig.~\ref{fig:am-detector-meas-true}.  In order to reject the current noise, these amplifier must be of the JFET type.  In the three detector scheme of Fig.~\ref{fig:am-detector-meas}\,B, it is convenient to use BJT amplifiers in the reference branches (A and B), and a JFET amplifier in the branch C\@.  The reason for this choice is the lower noise of the BJTs, which improves the measurement speed by reducing the minimum $m$ needed for a given sensitivity.

\section{AM noise in optical systems}\label{sec:am-optical-systems}
Equation~\req{eqn:am-def-alpha} also describes a quasi-perfect optical
signal, under the same hypotheses 1--4 of page \pageref{list:am-conditions}.  The voltage $v(t)$ is replaced with the electric field.
Yet, the preferred physical quantity used to describe the AM noise is the Relative
Intensity Noise (RIN), defined as
\begin{equation}
\mathrm{RIN} = S_\frac{\delta I}{I_0}(f)~~,
\end{equation}
that is, the power spectrum density of the normalized intensity fluctuation \begin{equation}
\frac{(\delta I)(t)}{I_0}=\frac{I(t)-I_0}{I_0}~~. 
\end{equation}
The RIN includes both fluctuation of power and the fluctuation of the power cross-section distribution.  If the cross-section distribution is constant in time, the optical intensity is proportional to power
\begin{equation}\frac{\delta I}{I_0}=\frac{\delta P}{P_0}~~.\end{equation}
In optical-fiber system, where the detector collects all the beam power,  the term RIN is improperly used for the relative power fluctuation.  
Reference \cite{su90apl} analyzes on the origin of RIN in semiconductor lasers, while References \cite{joindot92jpiii,obarski01josab} provide information on some topics of measurement.

In low-noise conditions, $\left|\delta I/I_0\right|\ll1$, and assuming that the cross-section distribution is constant, the power fluctuations are related to the fractional amplitude noise $\alpha$ by
\begin{equation}
\frac{\delta I}{I_0}=\frac{\delta P}{P_0}=2\alpha~~,
\end{equation}
thus
\begin{equation}
\mathrm{RIN}(f)=4S_\alpha(f)~~.
\end{equation}

Generally laser sources show a noise spectrum of the form
\begin{equation}
\mathrm{RIN}(f)=h_0+h_{-1}f^{-1}+h_{-2}f^{-2}~~,
\label{eqn:am-optical-rin}
\end{equation}
in which the flicker noise can be hidden by the random walk.
Additional fluctuations induced by the environment may be present.

\begin{figure}[t]
  \centering\namedgraphics{0.8}{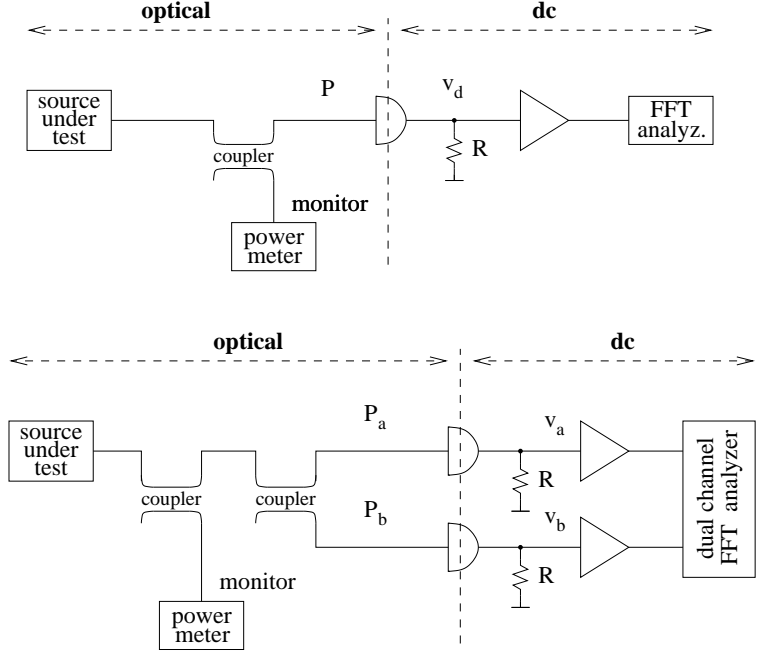}
\caption{RIN measurement in optical-fiber systems.  In a traditional
system, beam splitters are used instead of the couplers.}
\label{fig:am-optical}
\end{figure}

Figure \ref{fig:am-optical} shows two measurement schemes.  
The output signal of the photodetector is a current proportional to the photon flux.
Accordingly, the gain parameter is the detector responsivity $\rho$, defined by
\begin{equation}i=\rho P~~,\end{equation}
where 
\begin{equation}i=\frac{q\eta P}{h\nu}\end{equation}
is the photocurrent, thus 
\begin{equation}\rho=\frac{q\eta}{h\nu}~~.\end{equation}

The schemes of Fig.~\ref{fig:am-optical} are similar to Fig.~\ref{fig:am-basic-scheme} (single-channel) and Fig.~\ref{fig:am-correl-scheme} (cross-spectrum).   The dual-channel scheme is preferred because of the higher sensitivity, and because it makes possible to validate the measurement through the number of averaged spectra. 

Noise is easily analyzed with the methods shown in Section \ref{sec:am-amplifier}.  Yet in this case  the virtual-ground amplifier is often preferred, which differs slightly from the examples shown in Section \ref{sec:am-amplifier}.  A book \cite{graeme:photodiode-amplifiers} is entirely devoted to the special case of the photodiode amplifier.

\section{AM noise in microwave photonic systems}\label{sec:am-microwave-photonic}
Microwave and rf photonics is being progressively recognized as an emerging domain of technology \cite{chang02:rf-photonics,seeds:microwave-photonics}.  It is therefore natural to investigate in noise in these systems.

The power\footnote{In this section we use the subscript $\lambda$ for `light' and $\mu$ for `microwave'.} $P_\lambda(t)$ of the optical signal is sinusoidally
modulated in intensity at the microwave frequency $\nu_\mu$ is
\begin{equation}
   \label{eq:am-plambda-def}
   P_\lambda(t)=\overline{P}_\lambda\left(1+m\cos2\pi\nu_\mu t\right)~,
\end{equation}
where $m$ is the modulation index\footnote{We use the symbol $m$ for the modulation index, as in the general literature.  There is no ambiguity because the number of averages ($m$) is not used in this section.}.
Eq.~\req{eq:am-plambda-def} is similar to the traditional AM of radio broadcasting, but optical power is modulated instead of RF voltage.  In the presence of a distorted (nonlinear) modulation, we take the 
fundamental microwave frequency $\nu_0$.
The detector photocurrent is
\begin{equation}
       \label{eq:am-photoi-total}
       i(t)=\frac{q\eta}{h\nu_\lambda}\:
            \overline{P}_\lambda\left(1+m\cos2\pi\nu_\mu t\right)~,
\end{equation}
where $\eta$ the quantum efficiency of the photodetector.
The oscillation term $m\cos2\pi\nu_\mu t$ of Eq.~\req{eq:am-photoi-total} contributes to the microwave signal, the term ``1'' does not.
The microwave power fed into the load resistance $R_0$ is
$\overline{P}_\mu=R_0\smash{\overline{\tilde{\imath}^2}}$, hence
\begin{equation}
       \label{eq:am-pmu-gen}
       \overline{P}_\mu=\frac{1}{2}m^2R_0
             \left(\frac{q\eta}{h\nu_\lambda}\,\overline{P}_\lambda\right)^2~.
\end{equation}

The discrete nature of photons leads to the shot noise of power spectral
density $2qiR$ [W/Hz] at the detector output.  By virtue of Eq.~\req{eq:am-photoi-total},
\begin{equation}
       \label{eq:am-nsh-one-ch}
       N_s=2\frac{q^2\eta}{h\nu_\lambda}\:\overline{P}_\lambda R\qquad\text{(shot noise)}~.
\end{equation}
In addition, there is the equivalent input noise of the amplifier loaded by $R$, whose power spectrum is
\begin{equation}
       \label{eq:am-nth-def}
       N_t=FkT\qquad\text{(thermal noise and amplifier noise)}~,
\end{equation}
where $F$ is the noise figure of the amplifier, if any, at the output of the photodetector.
The white noise $N_s+N_t$ turns into a noise floor 
\begin{align}
S_{\alpha}=\frac{N_s+N_t}{\overline{P}_\mu}~.  
\end{align}
Using \req{eq:am-pmu-gen}, \req{eq:am-nsh-one-ch} and \req{eq:am-nth-def}, the floor is
\begin{equation}
       \label{eq:am-salpha0-one-arm}
       S_{\alpha}=\frac{2}{m^2}\left[
         2\frac{h\nu_\lambda}{\eta}\:\frac{1}{\overline{P}_\lambda}
         + \frac{FkT}{R}
            \left(\frac{h\nu_\lambda}{q\eta}\right)^2
            \left(\frac{1}{\overline{P}_\lambda}\right)^2
         \right]~.
\end{equation}
Interestingly, the noise floor is proportional to
$(\overline{P}_\lambda)^{-2}$ at low power, and to
$(\overline{P}_\lambda)^{-1}$ above the threshold power
\begin{equation}
       \label{eq:am-plambda-t}
       P_{\lambda,t}=\frac12\;\frac{FkT}{R}\;
          \frac{h\nu_\lambda}{q^2\eta}
\end{equation}
For example, taking $\nu_\lambda=193.4$ THz (wavelength $\lambda=1.55$
$\mu$m), $\eta=0.6$, $F=1$ (noise-free amplifier), and $m=1$, we get a
threshold power $P_{\lambda,t}=335$ $\mu$W, which sets the noise floor at
$5.1{\times}10^{-15}$ \unit{Hz^{-1}} ($-143$
\unit{dB/Hz}).

\begin{figure}[t]
  \centering\namedgraphics{0.8}{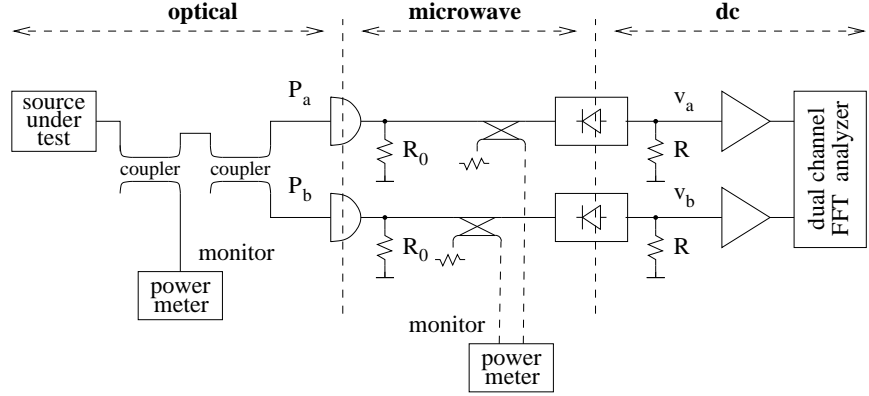}
\caption{Measurement of the microwave AM noise of a modulated light beam.}
\label{fig:am-mwave-photonic}
\end{figure}
Figure \ref{fig:am-mwave-photonic} shows the scheme of a correlation system for the measurement of the microwave AM noise.  It may be necessary to add a microwave amplifier at the output of each photodetector.
Eq.~\req{eq:am-salpha0-one-arm} holds for one arm of
Fig.~\ref{fig:am-mwave-photonic}.  As there are two independent arms, the noise
power is multiplied by two.  

Finally, it is to be pointed out that the results of this section concern only the white noise of the photodetector and of the microwave amplifier at the photodetector output.  
Experimental method and some data in the close-in microwave flickering of the high-speed photodetectors is available in Reference~\cite{rubiola06mtt-photodiodes}.
The noise of the microwave power detector and of its amplifier is still to be added, according to Section \ref{sec:am-detector-noise}.

\section{Calibration}\label{sec:am-calibration}
For small variations $\Delta P$ around a power $P_0$, the detector gain is replaced by the differential gain
\begin{equation}
k_d=\frac{d v_d}{d P}~~.
\label{eqn:am-meas-kd-def}
\end{equation}
which can be rewritten as
\begin{equation}
k_d=\frac{\Delta v_d}{\displaystyle\frac{\Delta P}{P_0}\,P_0}~~.
\label{eqn:am-meas-kd}
\end{equation}

Equations \req{eqn:am-salpha1-1channel}--\req{eqn:am-salpha2-1channel},
which are used to get $S_\alpha(f)$ from the spectrum $S_v(f)$
of the output voltage in single-channel measurements, rely upon the
knowledge of the calibration factor $k_dP_0$. The separate knowledge of
$k_d$ and $P_0$ is not necessary because only the product $k_dP_0$
enters in Eq.~\req{eqn:am-salpha1-1channel}--\req{eqn:am-salpha2-1channel}.
Therefore we can get $k_dP_0$ from
\begin{equation}
k_dP_0=\frac{\Delta v_d}{\Delta P/P_0}~~.
\label{eqn:am-meas-kdP0}
\end{equation}
This is a fortunate outcome for the following reasons
\begin{itemize}
\item A variable attenuator inserted in series to the oscillator under
  test sets a static $\delta P/P_0$ that is the same in all the
  circuit; this is a consequence of linearity.  For reference,
	\def\rrt{\rule[0ex]{0ex}{2.5ex}}
	\def\rrc{\rule[-1.5ex]{0ex}{4ex}}
	\def\rrb{\rule[-1.5ex]{0ex}{1.5ex}}
  \begin{center}
  \begin{tabular}{cc}
  step, dB	& $\Delta P/P_0$\rrb\\\hline
  0.1		& $2.33{\times}10^{-2}$\rrt\\
  0.5		& $0.122$\\
    1		& $0.259$
  \end{tabular}
  \end{center}
  \item A power ratio can be measured (or set) more accurately than an
    absolute power.
\end{itemize}
\begin{figure}[t]
  \centering\namedgraphics{0.8}{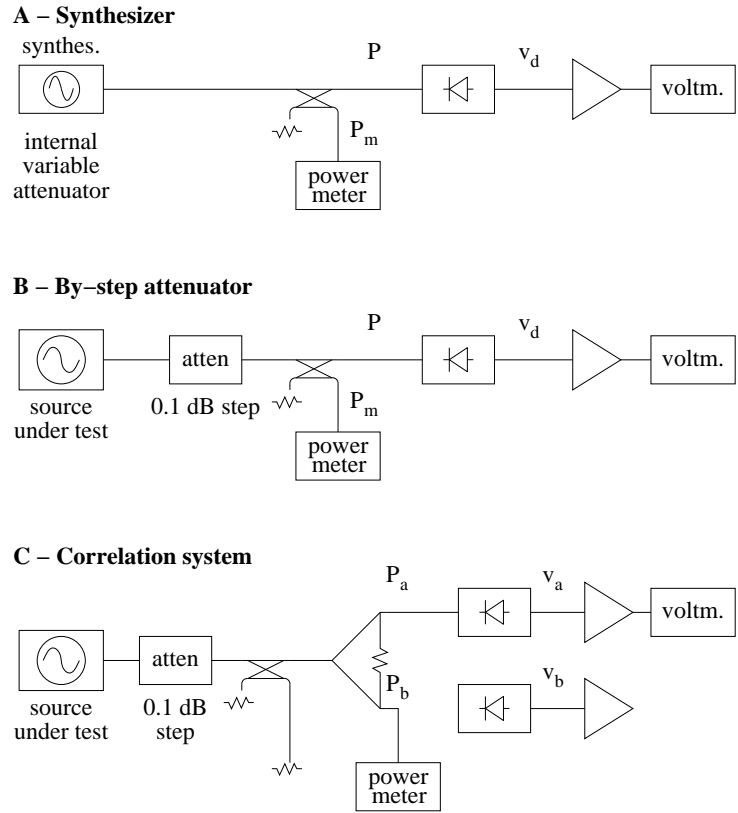}
\caption{Calibration schemes.}
\label{fig:am-calib-dc}
\end{figure}
Some strategies can be followed (Fig.~\ref{fig:am-calib-dc}), depending
on the available instrumentation.  In all cases it is recommended to
\begin{itemize}
\item make sure that the power detector works in the quadratic
  region (see Fig.~\ref{fig:am-detector-xfer}) by measuring the power
  at the detector input.
\item exploit the differential accuracy of the instruments that
  measure $\Delta P$ and $\Delta V$, instead of the absolute accuracy.
  Use the ``relative'' function if available, and do not change input
  range.
\item avoid plugging and unplugging connectors during the measurement.
  A directional coupler is needed not to disconnect the power detector
  for the measurement of $\Delta P$.
\end{itemize} 
In Fig.~\ref{fig:am-calib-dc}\,A, the internal variable attenuator of a
synthesizer is used to measure $k_dP_0$.  $\Delta P/P_0$ can be
measured with the power meter, or obtained from the calibration of the
synthesizer internal attenuator.  Some modern synthesizers have a
precise attenuator that exhibit a resolution of 0.1 or 0.01 dB\@.  In
Fig.~\ref{fig:am-calib-dc}\,B, a calibrated by-step attenuator is
inserted between the source under test and the power detector.
By-step attenuators can be accurate up to some 3--5 GHz.  Beyond, one
can use a multi-turn continuous attenuator and rely on the power
meter.  In the case of correlation measurements
(Fig.~\ref{fig:am-calib-dc}\,C), symmetry is exploited to measure $k_a$
and $k_b$ in a condition as close as possible to the final measurement
of $S_\alpha(f)$.  Of course, it holds that $\Delta P_a/P_a=\Delta P_b/P_b$.

\subsection{Alternate calibration method}\label{sec:am-alt-calibration}
\begin{figure}[t]
  \centering\namedgraphics{0.8}{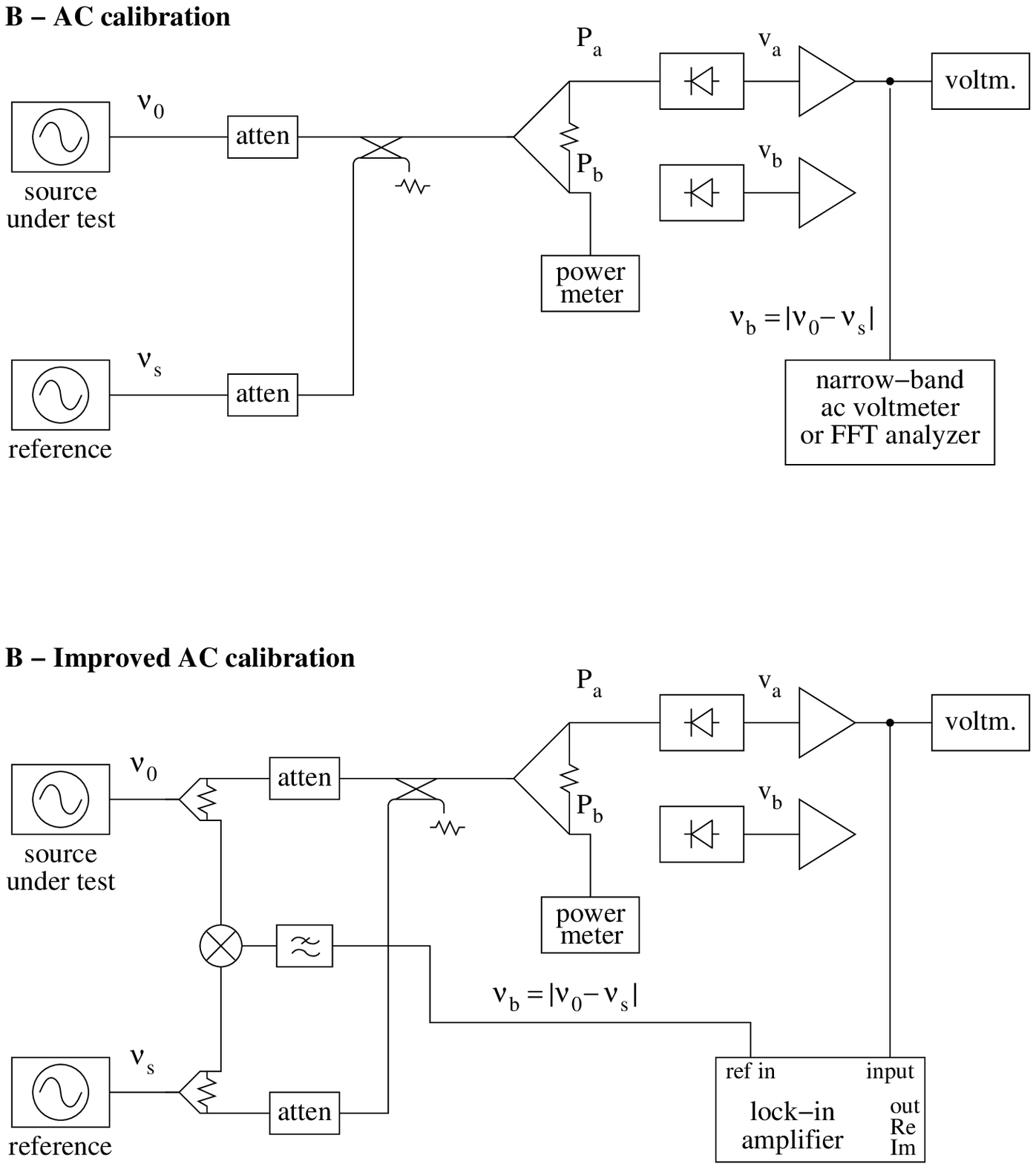}
\caption{Alternate calibration schemes.}
\label{fig:am-calib-ac}
\end{figure}
Another method to calibrate the power detector makes use of two synthesizers in the frequency region of interest, so that the beat note falls in the audio frequencies (Fig.~\ref{fig:am-calib-ac}).  This scheme is inspired to the two-tone method, chiefly used to measure the deviation of the detector from the ideal law $v_d=k_dP$ \cite{reinhardt95tmtt,walker04igarss}. 

Using $P=\smash{\frac{v^2}{R}}$, and denoting the carrier and the reference sideband with $v_0(t)=V_0\cos(2\pi\nu_0t)$ and $v_s(t)=V_s\cos(2\pi\nu_st)$, respectively, the detected signal is
\begin{align}
v_d(t) &= \frac{k_d}{R} \Bigl\{v_0(t)+v_s(t)\Bigr\}^2 * h_{lp}(t)~~.
\end{align} 
The low-pass function $h_{lp}$ keeps the dc and the beat note at the frequency $\nu_b=\nu_s-\nu_0$, and eliminates the $\nu_s+\nu_0$ terms.  Thus,
\begin{align}
v_d(t) &= \frac{k_d}{R} \Bigl\{\frac12V_0^2 + \frac12V_s^2 +2\frac12V_0V_s\cos\big[2\pi(\nu_s-\nu_0)t\big]\Bigr\}~~,
\end{align} 
which is split into the dc term
\begin{align}
\label{eqn:am-meas-alt-dc}
&\overline{v}_d = k_d\,\frac{V_0^2+V_s^2}{2R}
\intertext{and the beat-note term}
&\tilde{v}_d(t) = 2\,\frac{k_d}{R}\,\frac{V_0V_s}{2}\cos\big[2\pi(\nu_s-\nu_0)t\big]~~,
\intertext{hence}
\label{eqn:am-meas-alt-ac}
&\bigl(V_d\bigr)_\text{rms} = k_d\sqrt{2P_0P_s}~~.
\end{align} 

The dc term [Eq.~\req{eqn:am-meas-alt-dc}] makes it possible to measure $k_d$ from the contrast between $\overline{v}_1$, observed with the carrier alone, and $\overline{v}_2$, observed with both signals.  Thus, 
\begin{align}
k_d=\frac{\overline{v}_2-\overline{v}_1}{P_s}
\end{align}

Alternatively, the ac term [Eq.~\req{eqn:am-meas-alt-ac}] yields
\begin{align}
k_d=\frac{\bigl(V_d\bigr)_\text{rms}}{\sqrt{2P_0P_s}}
\end{align}
The latter is appealing because the assessment of $k_d$ relies only on ac measurements, which are free from offset and thermal drift.  On the other hand, the two-tone measurement does not provide the straight measurement of the product $k_dP_0$.

\section{Examples}\label{sec:am-examples}
\begin{figure}[t]
\centering%
\namedgraphics{0.8}{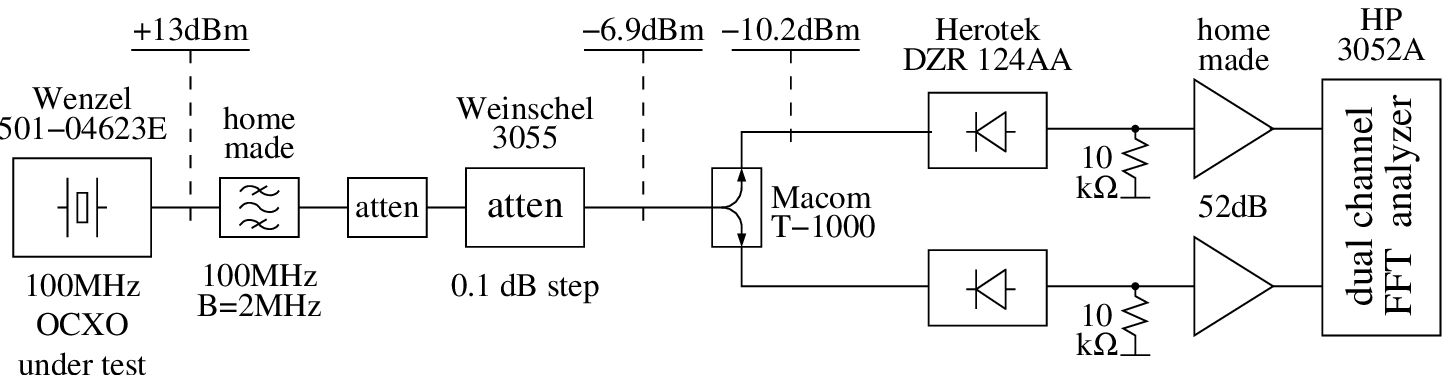}\\[2em]
\def\rrt{\rule[0ex]{0ex}{2.5ex}}
\def\rrc{\rule[-1.5ex]{0ex}{4ex}}
\def\rrb{\rule[-1.5ex]{0ex}{1.5ex}}
\centering
\begin{small}\begin{tabular}{l}\hline
\rrc Wenzel 501-04623E (s/n\ 3572-0214)\\\hline
\rrt $P_d=9.5$ $\mu$W ($-10.2$ dBm)\\
$k_d=1.31{\times}10^5$ A$^{-1}$ with dc ampli\\
$m=638$ averaged spectra\\
$1/\sqrt{2m}=35.7$ (15.5 dB)\\[1ex]
$h_{-1}=1.15{\times}10^{-12}$ Hz$^{-1}$ ($-129.4$ dB)\\
$\sigma_\alpha=4{\times}10^{-7}$\\\hline
\end{tabular}\end{small}\\[1ex]
\namedgraphics{0.7}{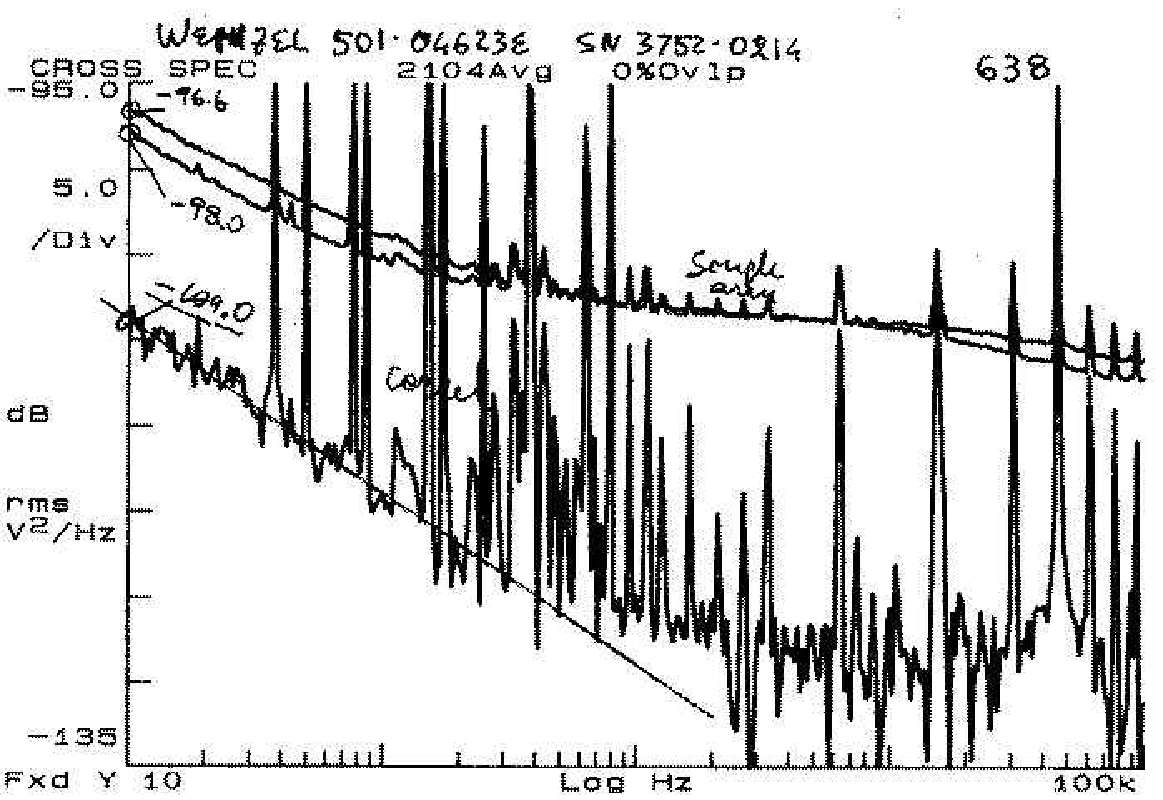}
\caption{Example of AM noise measurement\comment{ (Vol.~7 p.~28)}.}
\label{fig:am-v7p028b}
\end{figure}
Figure \ref{fig:am-v7p028b} show an 
example\comment{\footnote{Vol.~7 p.~28, (scheme at p.~18).}} of AM noise measurement.
The source under test is a 100 MHz quartz oscillator (Wenzel 501-04623E serial no.~3752-0214).

Calibration is done by changing the power $P_0=-10.2$~dBm by
${\pm}0.1$~dB.  There results $k_a=1.28{\times}10^5$~V/W and
$k_b=1.34{\times}10^5$~V/W, including the 52 dB amplifier (321~V/W and
336~V/W without amplification).  The system gain is therefore
$4k_ak_bP_aP_b=641$~\unit{V^2} (28.1~\unit{dBV^2}).

The cross spectrum of Fig.~\ref{fig:am-v7p028b} is
$S_{ba}=1.26{\times}10^{-11}$~\unit{V^2} ($-109$~\unit{dBV^2/Hz})
at 10 Hz, of the flicker type.  Averaging over $m=638$ spectra, the
single-channel noise is rejected by $\sqrt{2{\times}638}=35.7$
(15.5~dB).  The displayed flicker ($-109$ dB at 10 Hz) exceeds by only
3.8 dB the rejected single-channel noise.  A correction of a factor
0.58 ($-2.3$~dB) is therefore necessary\comment{In the first
  experiments (cf.\ my notebooks, Vol.~7 p.~28), I did not use the
  correction.}.  The corrected flicker is
$S_{ba}=7.4{\times}10^{-11}$~\unit{V^2} ($-101.3$~\unit{dBV^2/Hz})
extrapolated at 1 Hz.  The white noise can not be obtained
from Fig.~\ref{fig:am-v7p028b} because of the insufficient number
of averaged spectra.

As a consequence of the low amplitude noise of the oscillator, it is
possible to measure the noise of single channel, which includes detector and amplifier.  
Accounting for the gain (28.1~\unit{dBV^2}), the single-channel flicker noise of
Fig.~\ref{fig:am-v7p028b} at 1 Hz is
$S_\alpha(1\unit{Hz})=2.5{\times}10^{-12}$~\unit{Hz^{-1}}
($-116.1$~\unit{dB/Hz}) for one channel, and
$S_\alpha(1\unit{Hz})=3.4{\times}10^{-12}$~\unit{Hz^{-1}}
($-114.7$~\unit{dB/Hz}) for the other channel.

The AM flickering of the oscillator is
$S_\alpha(1\unit{Hz})=1.15{\times}10^{-13}$~\unit{Hz^{-1}}
($-129.4$ \unit{dB/Hz}), thus $h_{-1}=1.14{\times}10^{-13}$.  Using
the conversion formula of Tab.~\ref{tab:am-allan-variance} for flicker
noise, the Allan variance is $\sigma_\alpha^2=1.6{\times}10^{-13}$,
which indicates an amplitude stability
$\sigma_\alpha=4{\times}10^{-7}$,
independent of the measurement time $\tau$.

\begin{figure}[t]
\begin{minipage}[t]{0.495\textwidth}
\centering\textbf{\small A: Anritsu MG3690A synthesizer}\\[0.5ex]
\namedgraphics{0.45}{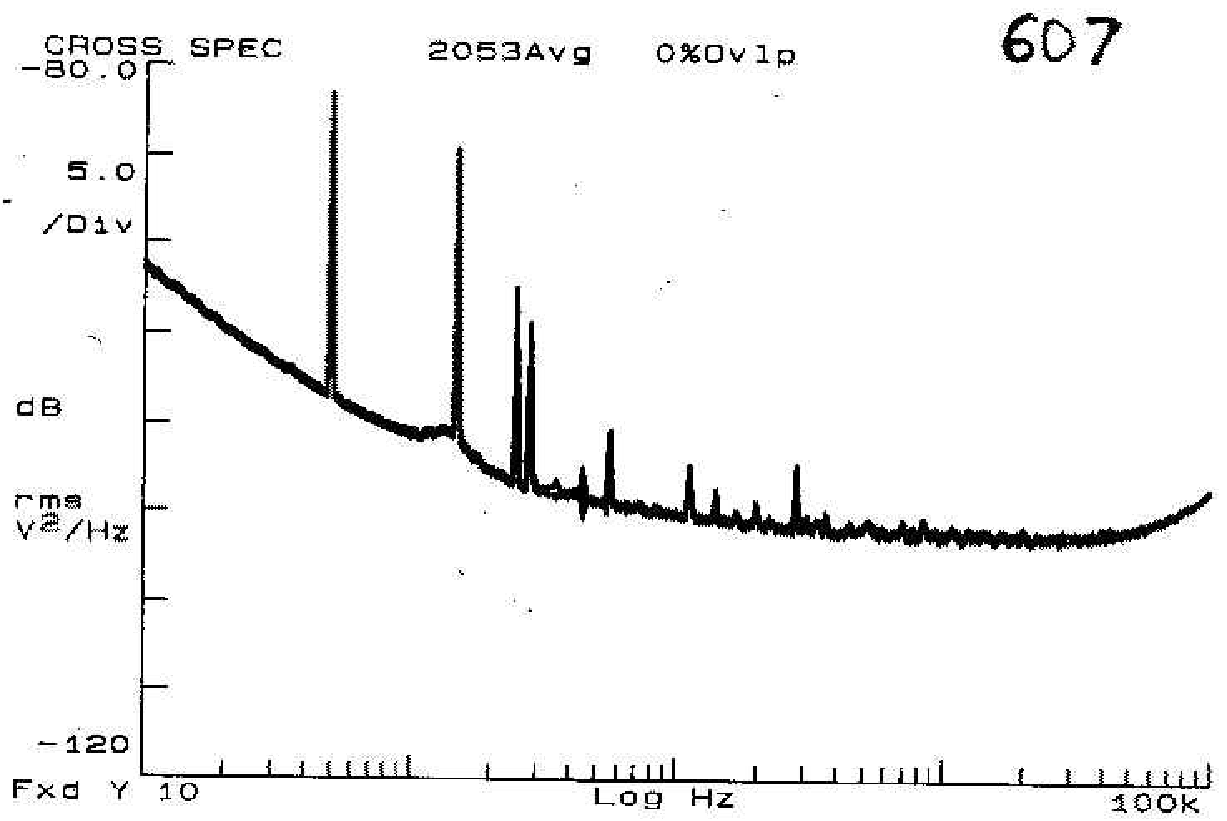}
\end{minipage}%
\begin{minipage}[t]{0.495\textwidth}
\centering\textbf{\small B: DRO Omega DRV9R192-105F}\\[0.5ex]
\namedgraphics{0.45}{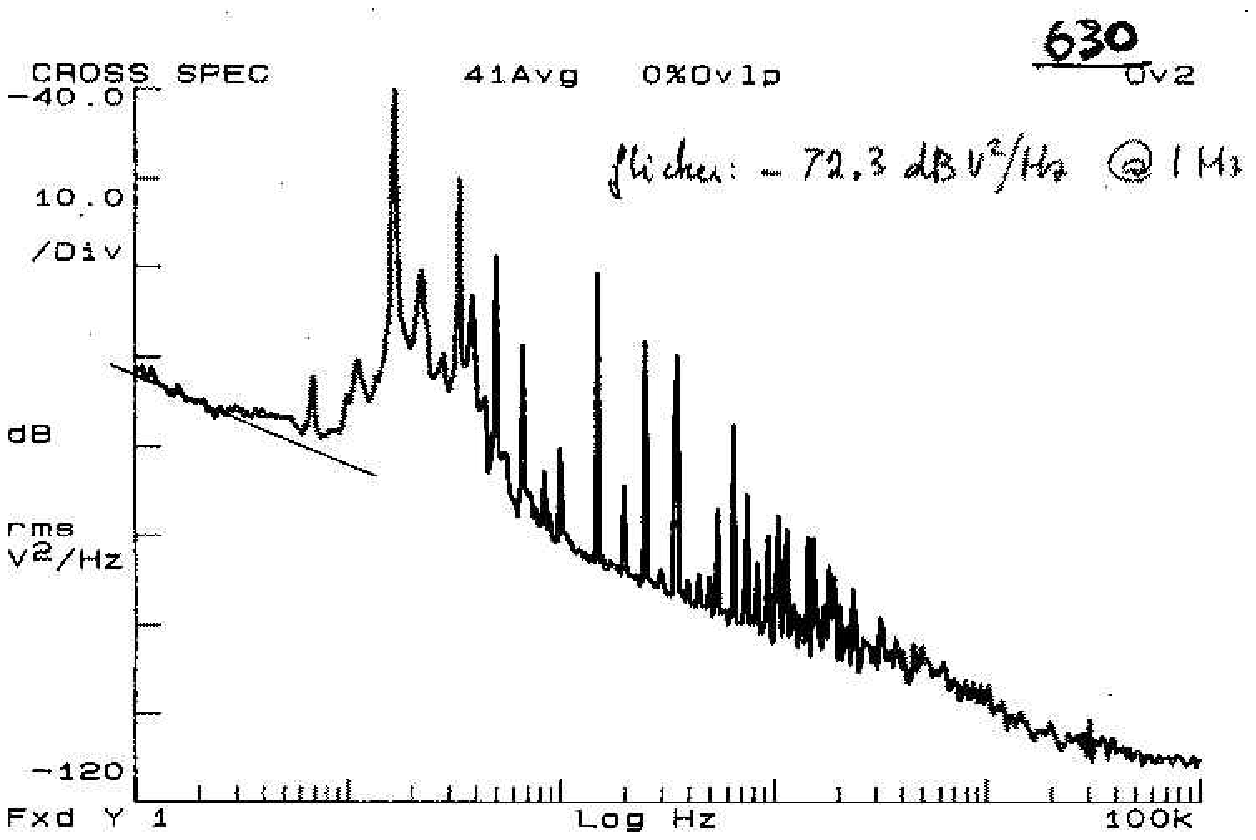}
\end{minipage}\\[1em]
\begin{minipage}[t]{0.495\textwidth}
\centering\textbf{\small C: HP 8662A synthesizer}\\[0.5ex]
\namedgraphics{0.45}{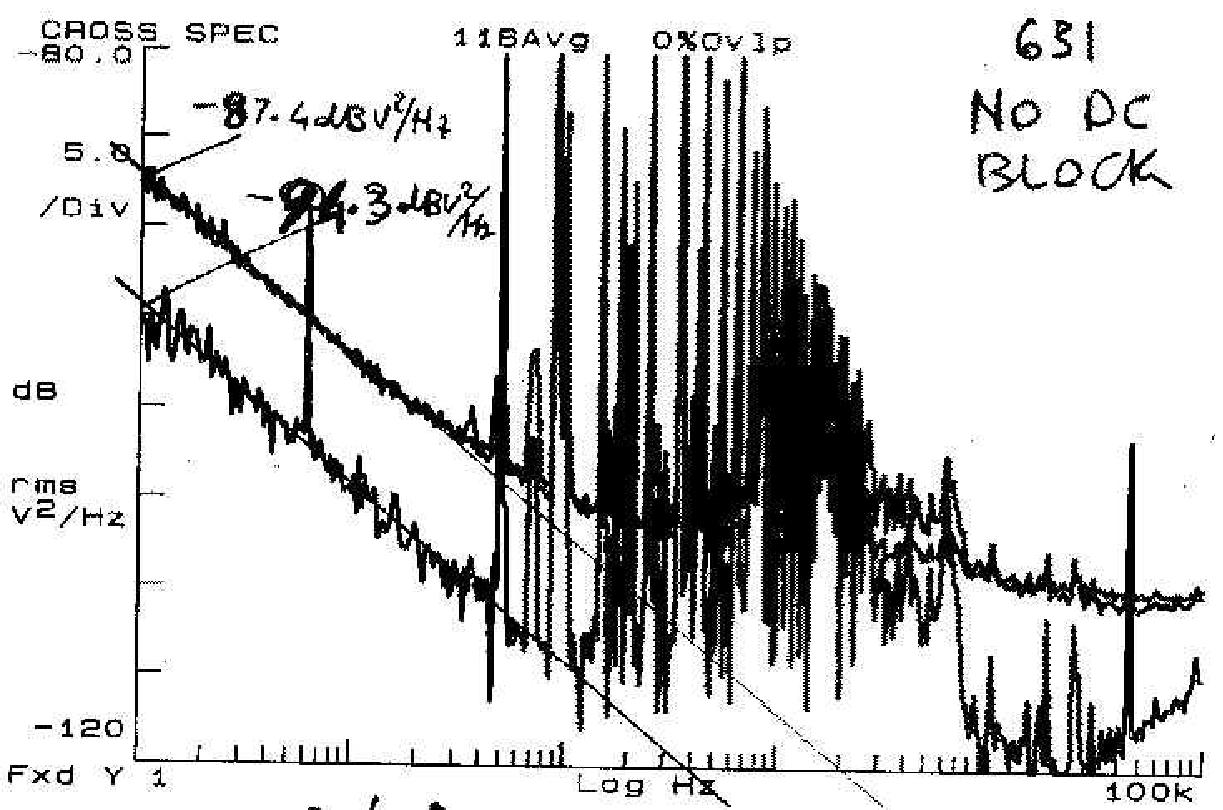}
\end{minipage}%
\begin{minipage}[t]{0.495\textwidth}
\centering\textbf{\small D: HP 8662A synthesizer}\\[0.5ex]
\namedgraphics{0.45}{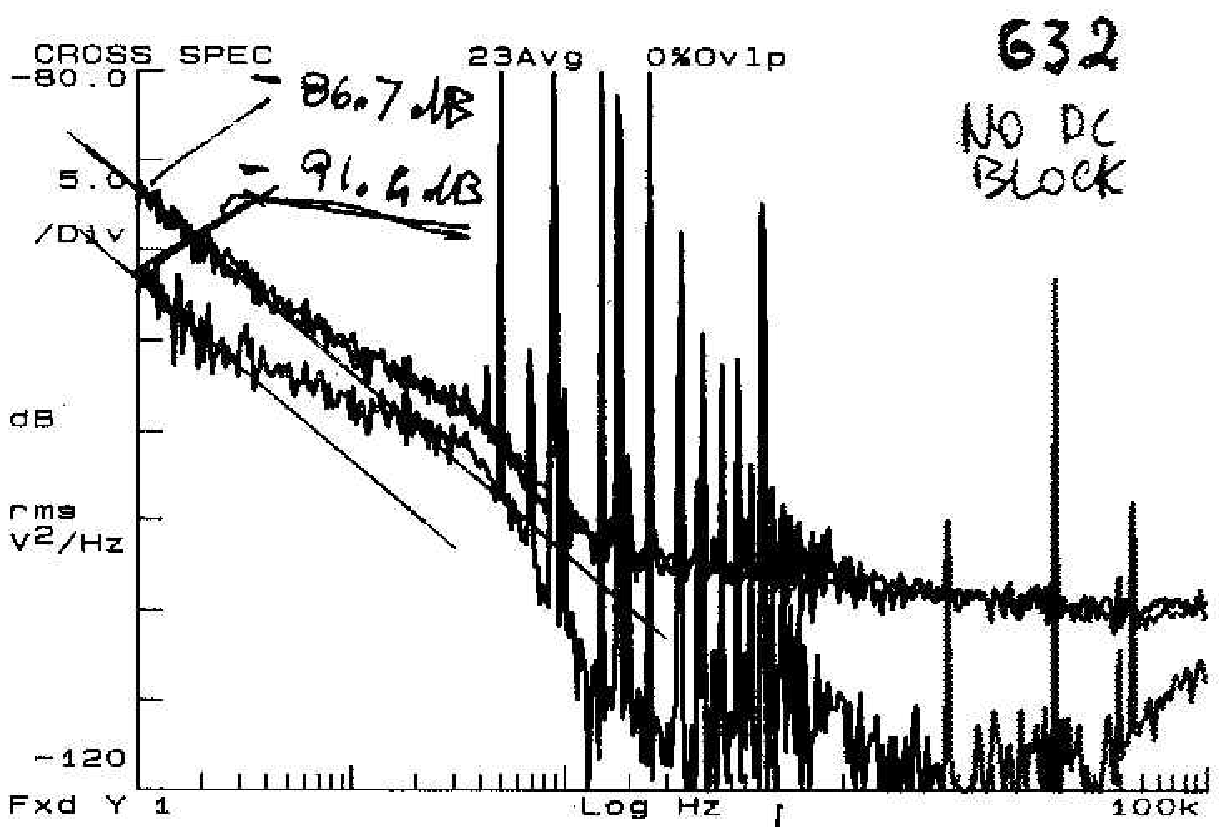}
\end{minipage}\\[1em]
\begin{minipage}[t]{0.495\textwidth}
\centering\textbf{\small E: Fluke 6160B synthesizer}\\[0.5ex]
\namedgraphics{0.45}{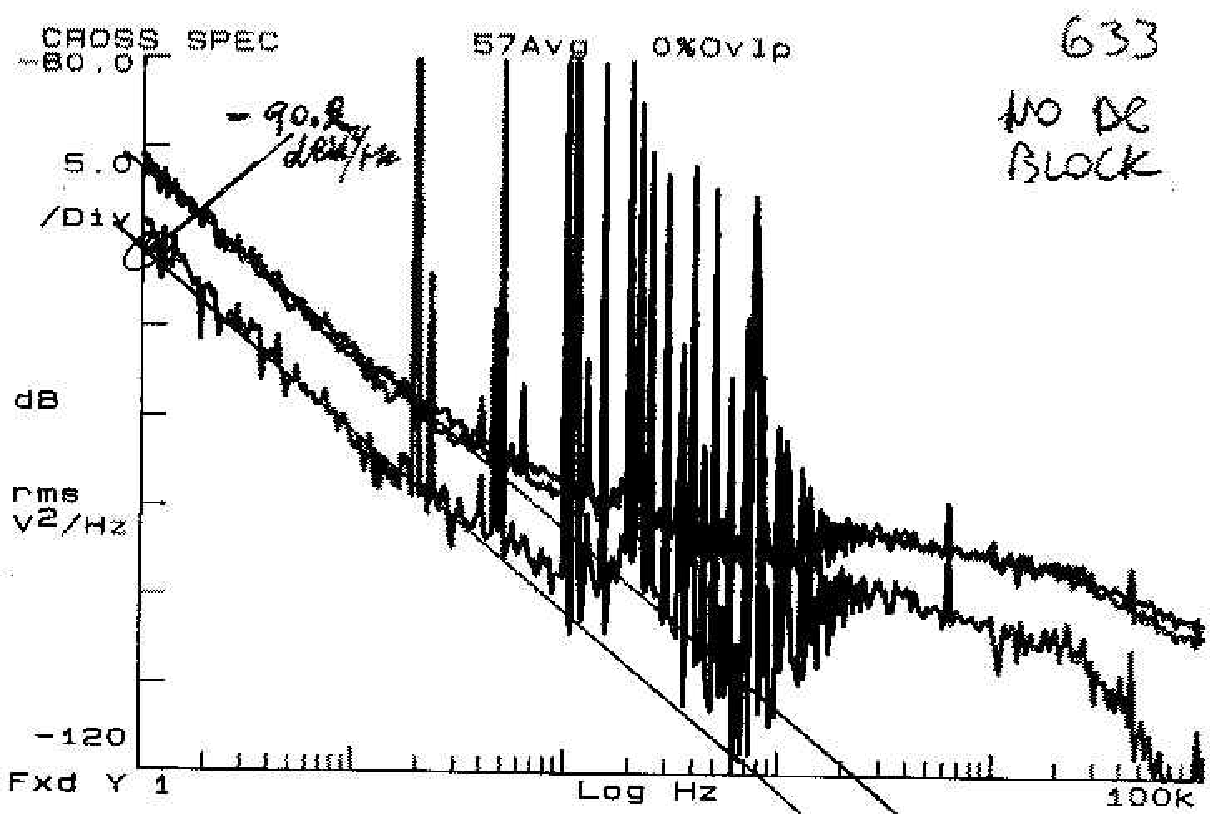}
\end{minipage}%
\begin{minipage}[t]{0.495\textwidth}
\centering\textbf{\small F: Racal Dana 9087B synthesizer}\\[0.5ex]
\namedgraphics{0.45}{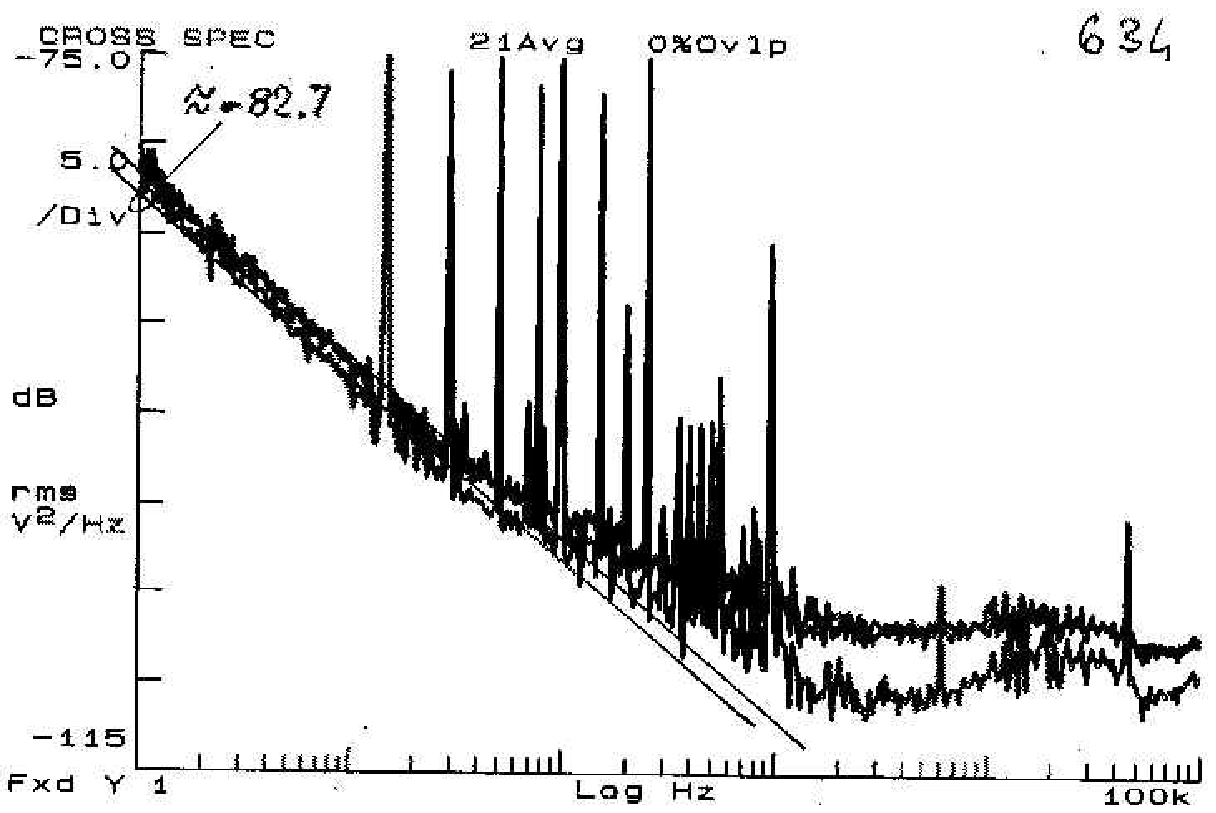}
\end{minipage}
\caption{Examples of AM noise measurement\comment{~(Vol.~6 p.~143)}.}
\label{fig:am-measured-spectra}
\end{figure}

\begin{figure}[t]
\centering\namedgraphics{0.64}{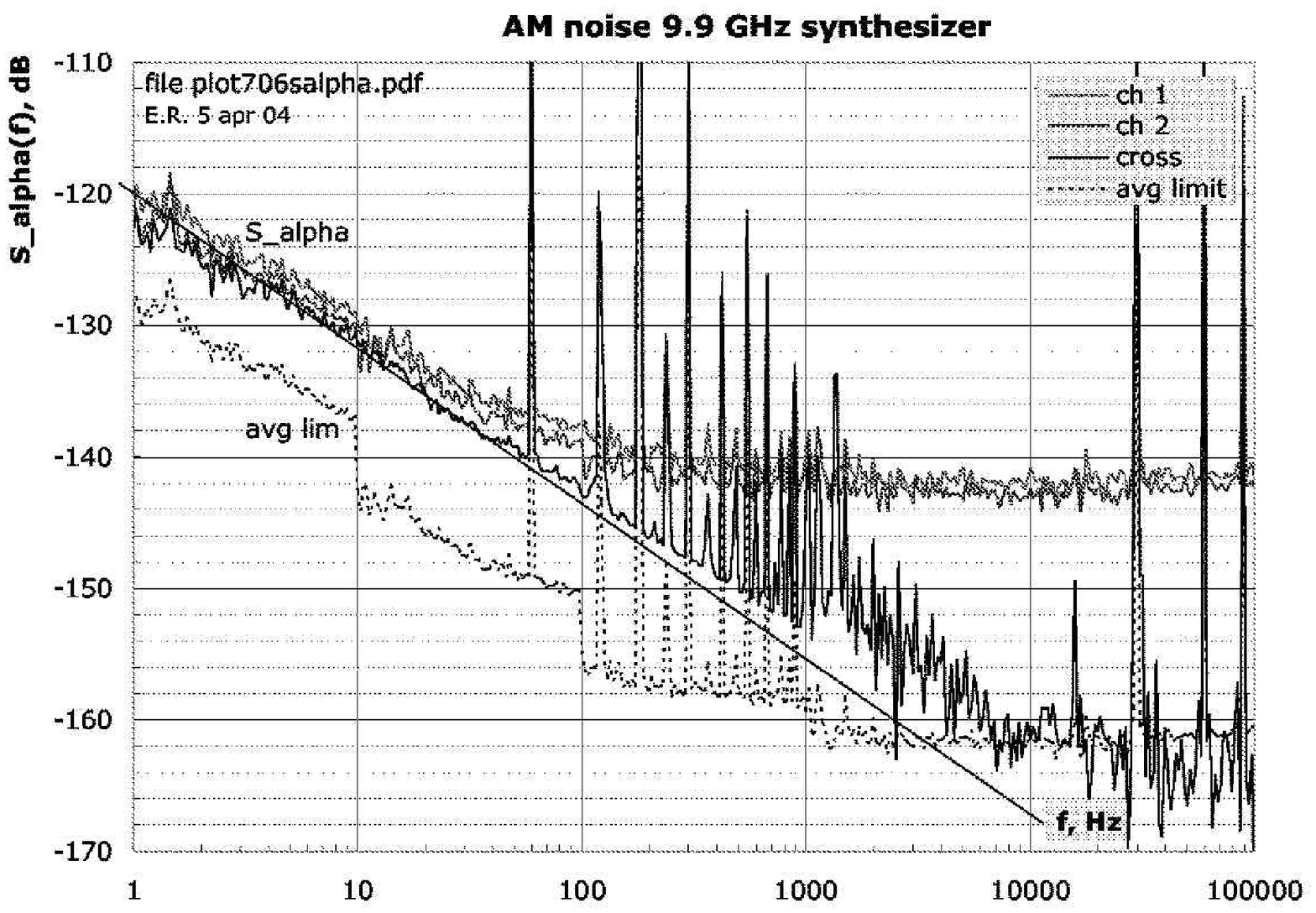}\\[1em]
\centering\namedgraphics{0.64}{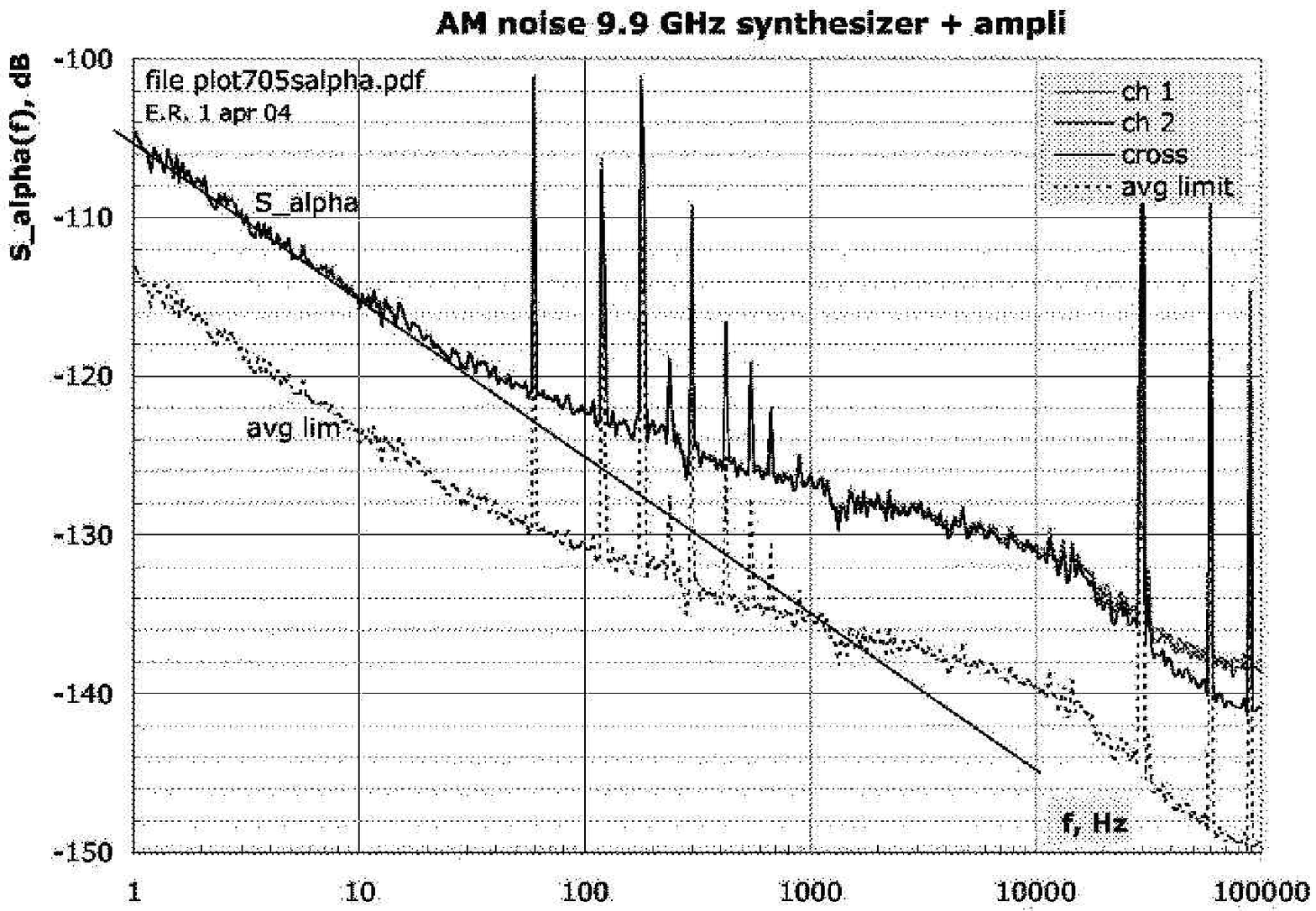}
\caption{Examples of AM noise measurement.\comment{~File am-plot706salpha and am-plot705salpha.}}
\label{fig:am-plot705salpha}
\end{figure}

Table~\ref{tab:am-sources} shows some examples of AM noise measurement.
The measured spectra are in Fig.~\ref{fig:am-v7p028b}, \ref{fig:am-measured-spectra}, and \ref{fig:am-plot705salpha}
\begin{table}[t]
\caption{AM noise of some sources.}
\label{tab:am-sources}
\centering
\begin{tabular}[t]{|l|c|c|l|}\hline\hline
\rule[-1ex]{0pt}{3.5ex}%
source & $h_{-1}$ & $\sigma_\alpha$ & notes \\\hline\hline
\rule{0pt}{2.5ex}%
Anritsu MG3690A & $2.5{\times}10^{-11}$ & $5.9{\times}10^{-6}$ 
  & Fig.~\ref{fig:am-measured-spectra}\,A\\
\rule[-1ex]{0pt}{1ex}%
synthesizer (10 GHz) & $-106.0$ dB & & \comment{Plot 607}
\comment{\\&&& VI-140/143/147}\\\hline
\rule{0pt}{2.5ex}%
Marconi & 1.1${\times}10^{-12}$ & $1.2{\times}10^{-6}$ & \\
\rule[-1ex]{0pt}{1ex}%
synthesizer (5 GHz) & $-119.6$ dB &  & \comment{VI-168}\\\hline
\rule{0pt}{2.5ex}%
Macom PLX 32-18 
  & 1.0${\times}10^{-12}$ & $1.2{\times}10^{-6}$ 
  & Fig.~\ref{fig:am-plot705salpha}\,A\\
\rule[-1ex]{0pt}{1ex}%
$0.1\rightarrow9.9$ GHz multiplier&$-120.0$ dB&&\comment{IX-133}\\
\hline
\rule{0pt}{2.5ex}%
Omega DRV9R192-105F & $8.1{\times}10^{-11}$ & $1.1{\times}10^{-6}$ 
  & Fig.~\ref{fig:am-measured-spectra}\,B\\
\rule[-1ex]{0pt}{1ex}%
9.2 GHz DRO & $-100.9$ dB &  & bump and junks
\comment{\\ &&& Plot 630, VII-13-14}\\\hline
\rule{0pt}{2.5ex}%
Narda DBP-0812N733 & $2.9{\times}10^{-11}$ & $6.3{\times}10^{-6}$ 
  & Fig.~\ref{fig:am-plot705salpha}\,A/B\\
\rule[-1ex]{0pt}{1ex}%
amplifier (9.9 GHz) & $-105.4$ dB &  & \comment{IX-125}\\\hline
\rule{0pt}{2.5ex}%
HP~8662A no.\,1 \comment{(CE122)} & $6.8{\times}10^{-13}$ & $9.7{\times}10^{-7}$ 
  & Fig.~\ref{fig:am-measured-spectra}\,C\\
\rule[-1ex]{0pt}{1ex}%
synthesizer (100 MHz) & $-121.7$ dB &  & junks
\comment{\\ &&& Plot 631, VII-17}\\\hline
\rule{0pt}{2.5ex}%
HP~8662A no.\,2 \comment{(CE125)} & $1.3{\times}10^{-12}$ & $1.4{\times}10^{-6}$ 
  & Fig.~\ref{fig:am-measured-spectra}\,D\\
\rule[-1ex]{0pt}{1ex}%
synthesizer (100 MHz) & $-118.8$ dB &  & junks
\comment{\\ &&& Plot 632, VII-18}\\\hline
\rule{0pt}{2.5ex}%
Fluke 6160B \comment{(CE124)} & $1.5{\times}10^{-12}$ & $1.5{\times}10^{-6}$ 
  & Fig.~\ref{fig:am-measured-spectra}\,E\\
\rule[-1ex]{0pt}{1ex}%
synthesizer           & $-118.3$ dB &  & junks
\comment{\\ &&& Plot 633, VII-19}\\\hline
\rule{0pt}{2.5ex}%
Racal Dana 9087B& $8.4{\times}10^{-12}$ & $3.4{\times}10^{-6}$ 
  & Fig.~\ref{fig:am-measured-spectra}\,F\\
\rule[-1ex]{0pt}{1ex}%
synthesizer (100 MHz) & $-110.8$ dB &  & junks
\comment{\\&&& Plot 634, VII-19}\\\hline
\rule{0pt}{2.5ex}%
Wenzel 500-02789D&$4.7{\times}10^{-12}$&$2.6{\times}10^{-6}$ 
  & Fig.~\ref{fig:am-v7p028b}\\
\rule[-1ex]{0pt}{1ex}%
100 MHz OCXO&$-113.3$ dB&  & \comment{Plot 638, VII-28}
\comment{\\ \#\,0627-9308 &&&} \\\hline
\rule{0pt}{2.5ex}%
Wenzel 501-04623E no.\,1&2.0${\times}10^{-13}$&$5.2{\times}10^{-7}$ 
  &\\
\rule[-1ex]{0pt}{1ex}%
100 MHz OCXO & $-127.1$ dB &  & \comment{VII-28}
\comment{\\ \#\,3752-0214 &&&}\\\hline
\rule{0pt}{2.5ex}%
Wenzel 501-04623E no.\,2& $1.3{\times}10^{-13}$ & $4.3{\times}10^{-7}$ 
  & \\
\rule[-1ex]{0pt}{1ex}%
100 MHz OCXO & $-128.8$ dB &  & \comment{VII-33}
\comment{\\ \#\,3753-0214 &&&}\\\hline\hline
\end{tabular}
\end{table}

All the experiments of Tab.~\ref{tab:am-sources} and Fig.~\ref{fig:am-v7p028b}, \ref{fig:am-measured-spectra}, and \ref{fig:am-plot705salpha} were done 
before thinking seriously about the design of the front-end amplifier (Section~\ref{sec:am-amplifier}), and before measuring the detector gain as a function of the load resistance (Table \ref{tab:am-conversion-gain}, and Figures \ref{fig:am-dzr124aa}--\ref{fig:am-dt8012}).  
The available low-noise amplifiers, designed for other purposes, turned out to be a bad choice, far from being optimized for this application.  Nonetheless, in all cases the observed cross spectrum is higher than the limit set by the average of two independent single-channel spectra.  In addition, the limit set by channel isolation is significantly lower than the observed cross spectrum.  These two facts indicate that the measured cross-spectrum is the true AM noise of the source.  Thus Table~\ref{tab:am-sources} is an accurate database for a few specific cases.  Of course, Table~\ref{tab:am-sources} also provides the order of magnitude for the AM noise of other synthesizers and oscillators employing similar technology.
On the other hand, the data of Table~\ref{tab:am-sources} do not provide information on the detector noise. 

The amplifier used in almost all the experiments is the ``super-low-noise amplifier'' proposed in the data sheet of the MAT03 \cite[Fig.~3]{pmi:mat03spec}, which is matched PNP transistor pair.  For reference, the NPN version of this amplifier is discussed in \cite{franco:operational-amplifiers}.  The input differential stage of this amplifier consists of three differential pairs connected in parallel, so that the voltage noise is the noise of a pair divided by $\sqrt{3}$.  Yet the current noise is multiplied by $\sqrt{3}$.  As a consequence, the amplifier is noise-matched to an impedance of some 200 \ohm\ for flicker noise, and to some 30 \ohm\ for white noise, which is too low for our purposes.
The second version of the MAT03 amplifier, designed after the described experiments, was optimized for the lowest flicker when connected to a 50 \ohm\ source \cite{rubiola04rsi-amplifier}.  This amplifier, now routinely employed for the measurement of phase noise, makes use \emph{one} MAT03 instead of three.  In two cases (Fig.~\ref{fig:am-plot705salpha}) a different amplifier was used, based on the OP37 operational amplifier loaded to an input resistance of some 1 k\ohm.  Interestingly, in the operating conditions of AM noise measurements, the OP37 outperforms the more sophisticated MAT03.

\section{Final remarks}
\paragraph{True quadratic detection vs.\ peak detection.}
Beyond a threshold power, a power detector leaves the quadratic operations and
works as a peak detector.  The peak detection is the same operation mode of the
old good detectors for AM broadcasting (which is actually an \emph{envelope} modulation).
This operation mode exhibits higher gain, hence it could be advantageous
for the measurement of low-noise signals.    The answer may depend on the diode type, Schottky or tunnel.  The strong recommendation to use the diode in the quadratic region might be wrong.

\paragraph{Trans-resistance amplifiers.}
In principle, the power detector can be used as a power-to-current converter (instead of as a power-to-voltage) converter, and connected to a trans-resistance amplifier.  
The advantage is that the resistor at the detector output, which is a relevant source of noise in voltage-mode measurements, is not present.   
This choice, suggested in \cite{burrus63mtt}, is never found in the technical literature accompanying the detectors.

\paragraph{Cryogenic environment.}
In principle, the tunnel diode should work at cryogenic temperatures.
Yet, the laboratory could be much less smooth than the theory.

\def\bibfile#1{/Users/rubiola/Documents/work/bib/#1}
\addcontentsline{toc}{section}{References}
\bibliographystyle{amsalpha}
\bibliography{\bibfile{ref-short},%
              \bibfile{references},%
              \bibfile{rubiola}}

\end{document}